\pdfoutput=1
\documentclass[conference]{IEEEtran}

\usepackage{mathptmx} 
\usepackage{fancyhdr}
\usepackage[normalem]{ulem}
\usepackage[hyphens]{url}
\usepackage[sort,nocompress]{cite}
\usepackage[final]{microtype}
\usepackage[keeplastbox]{flushend}
\usepackage[bookmarks=true,breaklinks=true,letterpaper=true,colorlinks,linkcolor=black,citecolor=blue,urlcolor=black]{hyperref}
\usepackage{xspace}

\newcommand{\ignore}[1]{}
\usepackage[pass]{geometry}
\usepackage{color}
\usepackage{soul}
\usepackage{siunitx}
\usepackage{tikz}
\usepackage{graphicx}
\usepackage{multirow}
\usepackage{array}
\newcolumntype{P}[1]{>{\centering\arraybackslash}p{#1}}
\usepackage{dblfloatfix}
\usepackage{float}
\usepackage[font={small,bf}]{caption}

\usepackage[inline]{enumitem}
\newcommand*\circled[1]{\tikz[baseline=(char.base)]{%
            \node[shape=circle,fill=gray!20,draw,inner sep=1.5pt] (char) {#1};}}

\definecolor{aquamarine}{rgb}{0.5, 1.0, 0.83}
\definecolor{ashgrey}{rgb}{0.7, 0.75, 0.71}
\definecolor{atomictangerine}{rgb}{1.0, 0.6, 0.4}
\definecolor{babyblue}{rgb}{0.54, 0.81, 0.94}
\definecolor{fluorescentyellow}{rgb}{0.8, 1.0, 0.0}
\definecolor{lavender(floral)}{rgb}{0.71, 0.49, 0.86}
\definecolor{mauve}{rgb}{0.88, 0.69, 1.0}

\fancypagestyle{firstpage}{
  \fancyhf{}
  
  \fancyfoot[C]{\thepage}
}

\title{\huge{Architecting Optically-Controlled Phase Change Memory}}
\author{\IEEEauthorblockN{Aditya Narayan\IEEEauthorrefmark{1},
Yvain Thonnart\IEEEauthorrefmark{2},
Pascal Vivet\IEEEauthorrefmark{2},
Ayse K. Coskun\IEEEauthorrefmark{1} and
Ajay Joshi\IEEEauthorrefmark{1}}
\IEEEauthorblockA{\IEEEauthorrefmark{1}Boston University, Boston, MA 02215, USA;
Email: \{adityan, acoskun, joshi\}@bu.edu}
\IEEEauthorblockA{\IEEEauthorrefmark{2}Univ. Grenoble Alpes, CEA, List, Grenoble, France;
Email: \{yvain.thonnart, pascal.vivet\}@cea.fr}
}

\newcommand{\optmemory}{OPCM\xspace}
\newcommand{\opcmsystem}{COSMOS\xspace}
\newcommand{\eoe}{E-O-E control\xspace}

\newcommand{\NEW}{\color{black}}

\usepackage{marginnote}
\usepackage[caption=false]{subfig}

\begin{document}
\newpage

\maketitle
\pagenumbering{arabic}

\thispagestyle{firstpage}
\pagestyle{plain}
\setcounter{page}{1}


\begin{abstract}

Phase Change Memory (PCM) is an attractive candidate for main memory as it offers non-volatility and zero leakage power, while providing higher cell densities, longer data retention time, and higher capacity scaling compared to DRAM.
In PCM, data is stored in the crystalline or amorphous state of the phase change material.
The typical electrically-controlled PCM (EPCM), however, suffers from longer write latency and higher write energy compared to DRAM and limited multi-level cell (MLC) capacities.
These challenges limit the performance of data-intensive applications running on computing systems with EPCMs.

Recently, researchers demonstrated optically-controlled PCM (OPCM) cells, with support for $5$ $bits/cell$ in contrast to 
$2$ $bits/cell$ in EPCM.
These \optmemory cells can be accessed directly with optical signals that are multiplexed in high-bandwidth-density silicon-photonic links.
The higher MLC capacity in \optmemory and the direct cell access using optical signals enable an increased read/write throughput and lower energy per access than EPCM.
However, due to the direct cell access using optical signals, \optmemory systems cannot be designed using conventional memory architecture.
We need a complete redesign of the memory architecture that is tailored to the properties of \optmemory technology.

This paper presents the design of a unified network and main memory system called  \opcmsystem that combines \optmemory and silicon-photonic links to achieve high memory throughput.
\opcmsystem is composed of a hierarchical multi-banked \optmemory array with novel read and write access protocols.
\opcmsystem uses an Electrical-Optical-Electrical (E-O-E) control unit to map standard DRAM read/write commands (sent in electrical domain) from the memory controller on to optical signals that access the \optmemory cells.
Our evaluation of a 2.5D-integrated system containing a processor and \opcmsystem demonstrates $2.14\times$ average speedup across graph and HPC workloads compared to an EPCM system.
\opcmsystem consumes $3.8\times$ lower read energy-per-bit and $5.97\times$ lower write energy-per-bit compared to EPCM.
\opcmsystem is the first non-volatile memory that provides comparable performance and energy consumption as DDR4 in addition to increased bit density, higher area efficiency and improved scalability.
\end{abstract}

\section{Introduction}
\label{sec:introduction}

Today's data-driven applications that use graph processing~\cite{malewicz2010pregel, lowGraphLab, salihoglu2013gps, gonzalez2012powergraph}, machine learning~\cite{brown2020language, woodhouse2016big, 8740962}, or privacy-preserving paradigms~\cite{shi2016edge, acar2018survey, cheon2017homomorphic} demand memory sizes on the order of hundreds of $GBs$ and bandwidths on the order of $TB/s$.
The widely-used main memory technology, DRAM, is facing critical technology scaling challenges and fails to meet the increasing bandwidth and capacity demands of these data-driven applications~\cite{kim2010capacitors,kim2018future,mutlu2013memory,kang2014co,xu2015overcoming, lefurgy2003energy}.
Phase Change Memory (PCM) is emerging as a class of non-volatile memory (NVM) that is a promising alternative to DRAM~\cite{qureshi2009scalable, lee2011energy, lee2011characterizing, jia2016dynamic, ramos2011page, kim2019ll}.
PCMs outperform other NVM candidates owing to their higher reliability, increased bit density, and better write endurance~\cite{bedeschi2008multi, burr2010phase, wuttig2017phase, nirschl2007write}.
In PCMs, data is stored in the state of the phase change material, i.e., crystalline (logic $1$) or amorphous (logic $0$)~\cite{ovshinsky1968reversible, wuttig2007phase}.
A SET operation triggers a transition to crystalline state, and a RESET operation triggers a transition to amorphous state.
PCMs also enable multi-level cells (MLC) using the partially crystalline states.
Higher MLC capacity enables increased bit density ($bits/mm^2$).

PCM cells are typically controlled electrically (we refer to them as EPCM cells), where different PCM states have distinct resistance values.
Main memory systems using EPCM cells are designed using the same microarchitecture and read/write access protocol as DRAM systems~\cite{lee2009architecting, thakkar2017dyphase}.
EPCM cells are SET or RESET by passing the corresponding current through the phase change material (via the bitline) to trigger the desired state transition.
The state of the EPCM cells is read out by passing a read current and measuring the voltage on the bitline.
EPCM systems, however, are limited to $2$ $bits/cell$~\cite{bedeschi2008multi, nirschl2007write, cabrini2009voltage} due to resistance drift over time, have $3-4\times$ higher write latency than DRAM leading to lower performance~\cite{lee2009architecting,arjomand2016boosting}, consume high power due to the need for large on-chip charge pumps~\cite{jiang2014low, palumbo2010charge, wong2012comparative}, and have lower lifetime than DRAM due to faster cell wearout~\cite{qureshi2009enhancing}. 

Recent advances in device research have demonstrated optically-controlled PCM cells (we refer to them as \optmemory cells)~\cite{rios2015integrated, feldmann2017calculating, feldmann2019integrated, carrillo2019behavioral}.
\optmemory cells exhibit higher MLC capacity than EPCM cells (up to $5$ $bits/cell$~\cite{li2019fast}).
Moreover, high-bandwidth-density silicon-photonic links\cite{sun2015single,TeraPHY}, which are being developed for processor-to-memory communication, can directly access these \optmemory cells, thereby yielding higher throughput and lower energy-per-access than EPCM.
These two factors make \optmemory a more attractive candidate for main memory than EPCM.

\begin{figure}[t]
    \centering
    \includegraphics[width=0.9\columnwidth]{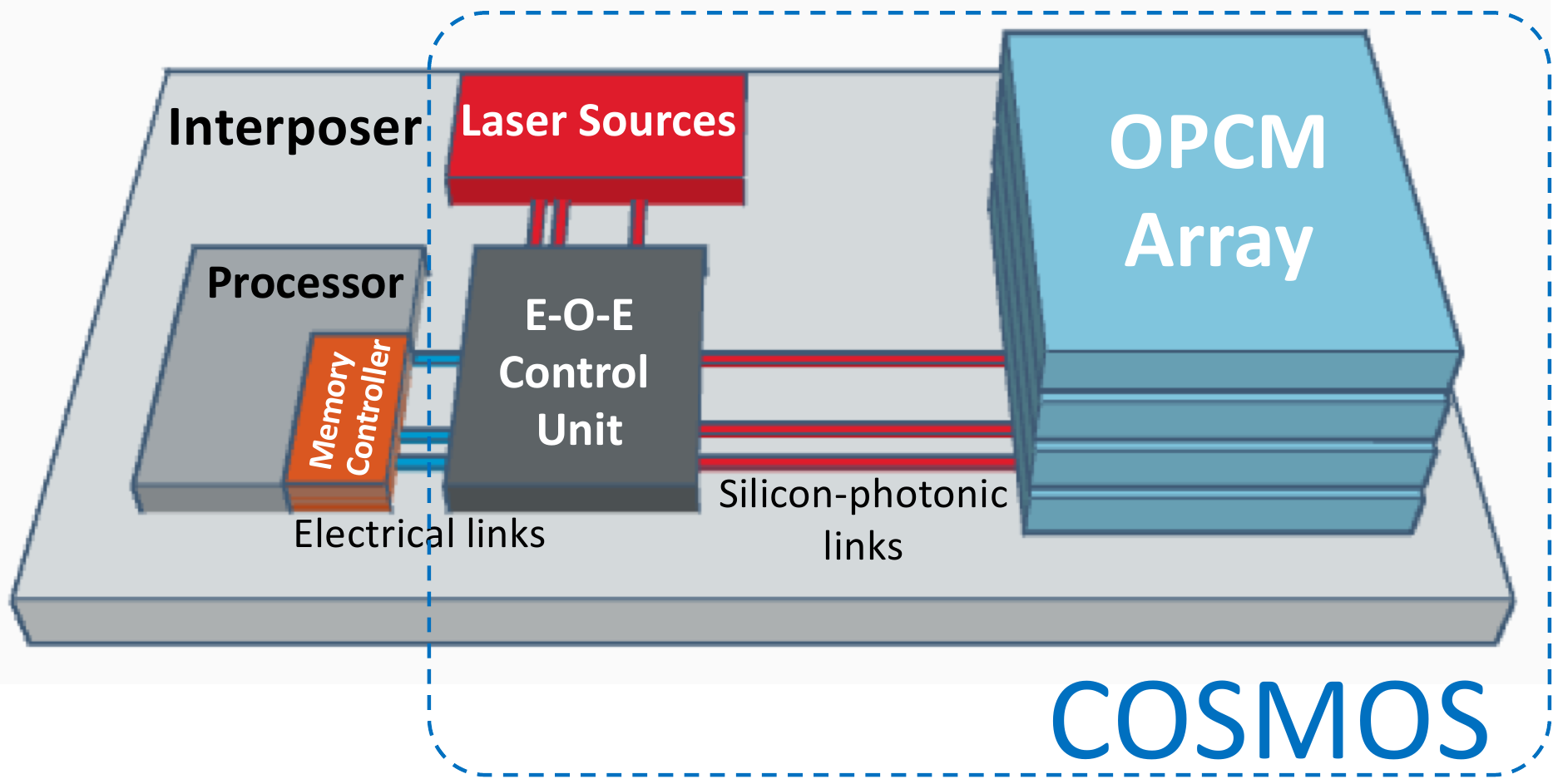}
    \caption[Caption for LOF]{Overview of a 2.5D-integrated computing system with \optmemory array stack as the main memory, \eoe unit chiplet, processor chiplet, and laser sources chiplet.\protect\footnotemark}
    \label{fig:sys_overview}
    \vspace{-0.2in}
\end{figure}

Since the optical signals in silicon-photonic links directly access the \optmemory cells, the traditional row-buffer based memory microarchitecture and the read/write access protocol encounter critical design challenges when adapted for \optmemory.
We need a complete redesign of the memory microarchitecture and a novel access protocol that is tailored to the \optmemory cell technology.
\textbf{In this paper, we propose a \underline{CO}m{\-}bined \underline{S}ystem of Optical Phase Change \underline{M}emory and \underline{O}p{\-}tical Link\underline{S}, \opcmsystem, which integrates the \optmemory technology and the silicon-photonic link technology, thereby providing seamless high-bandwidth access from the processor to a high-density memory.}
Figure~\ref{fig:sys_overview} shows a computing system with \opcmsystem.
\opcmsystem includes a hierarchical multi-banked \optmemory array, \eoe unit, silicon-photonic links, and laser sources. 
The multi-banked \optmemory array uses 3D optical integration to stack multiple banks vertically, with 1 bank/layer.
The cells in the \optmemory array are directly accessed using silicon-photonic links that carry optical signals, thereby eliminating the need for electrical-optical (E-O) and optical-electrical (O-E) conversion in the \optmemory array.
These optical signals are generated by an \eoe unit that serves as an intermediary between the memory controller (MC) in the processor and the \optmemory array.
This \eoe unit is responsible for mapping the standard DRAM protocol commands sent by the MC onto optical signals, and then sending these optical signals to the \optmemory array.
%
The major contributions of our work are as follows:
\begin{enumerate}[leftmargin=\parindent,align=left,labelwidth=\parindent,labelsep=0pt]
\item We architect the \opcmsystem, which consists of a hierarchical multi-banked \optmemory array, where the cells are accessed directly using optical signals in silicon-photonic links.
The \optmemory array design combines wavelength-division-multiplexing (WDM) and mode-division-multiplexing (MD\-M) properties of optical signals to deliver high memory bandwidth.
Moreover, the \optmemory array contains only passive optical elements and does not consume power, thus providing cost and efficiency advantages.
\item We propose a novel mechanism for read and write operation of cache lines in \opcmsystem.
A cache line is interleaved across multiple banks in the \optmemory array to enable high-throughput access.
The write data is encoded in the intensity of optical signals that uniquely address the \optmemory cell.
The readout of an \optmemory cell uses a 3-step operation that measures the attenuation of the optical signal transmitted through the cell, where the attenuation corresponds to a predetermined bit pattern.
Since the read operation is destructive, we design an opportunistic writeback operation of the read data to restore the \optmemory cell state.
\item  We design an \eoe unit to interface \opcmsystem with the processor. This \eoe unit receives standard DRAM commands from the processor, and converts them into the \optmemory-specific address, data, and control signals that are mapped onto optical signals.
These optical signals are then used to read/write data from/to the \optmemory array.
The responses from the \optmemory array are converted by the \eoe unit back into standard DRAM protocol commands that are sent to the processor. 
\end{enumerate}

\noindent Evaluation of a 2.5D system with a multi-core processor and \opcmsystem demonstrates $2.15\times$ higher  write throughput and $2.09\times$ higher read  throughput compared to an equivalent system with EPCM.
For graph and high performance computing (HPC) workloads, when compared to EPCM, \opcmsystem has $2.14\times$ better performance, $3.8\times$ lower read energy-per-bit, and $5.97\times$ lower  write energy-per-bit.
Moreover, \opcmsystem provides a scalable and non-volatile alternative to DDR4 DRAM systems, with $5.6\%$ higher performance and similar energy consumption for read and write accesses.
With DRAM technology undergoing critical scaling challenges, \opcmsystem presents the first non-volatile main memory system with improved scalability, increased bit density, high area efficiency, and comparable performance and energy consumption as DDR4 DRAM.

\footnotetext{\opcmsystem-based computing system is agnostic of the integration technology. 
However, 3D-integrated systems raise thermal concerns and 2D systems result in large system footprints and communication overheads.}
    
\ignore{The rest of the paper is organized as follows: 
Section~\ref{sec:background} provides a background on the operation of \optmemory cells.
Section~\ref{sec:motivation} explains why today's main memory architecture cannot be directly adapted for \optmemory.
Section~\ref{sec:opcm_arch} describes the \optmemory array microarchitecture in \opcmsystem, and the read and write access protocol.
Section~\ref{sec:EOE_unit} provides details of the \eoe unit.
We describe our evaluation methodology in Section~\ref{sec:eval_meth} and present our evaluation results in Section~\ref{sec:revision_results}.}
\section{Background}
\label{sec:background}

In this section, we discuss the basic operation of an \optmemory cell along with its properties, and the silicon-photonic links that enable optical signals to directly access the \optmemory cells.
\subsection{\optmemory Cell}
\begin{figure}[t]
    \centering
        \subfloat{
            \label{fig:OPCM_3D}
            \includegraphics[height=1.1in]{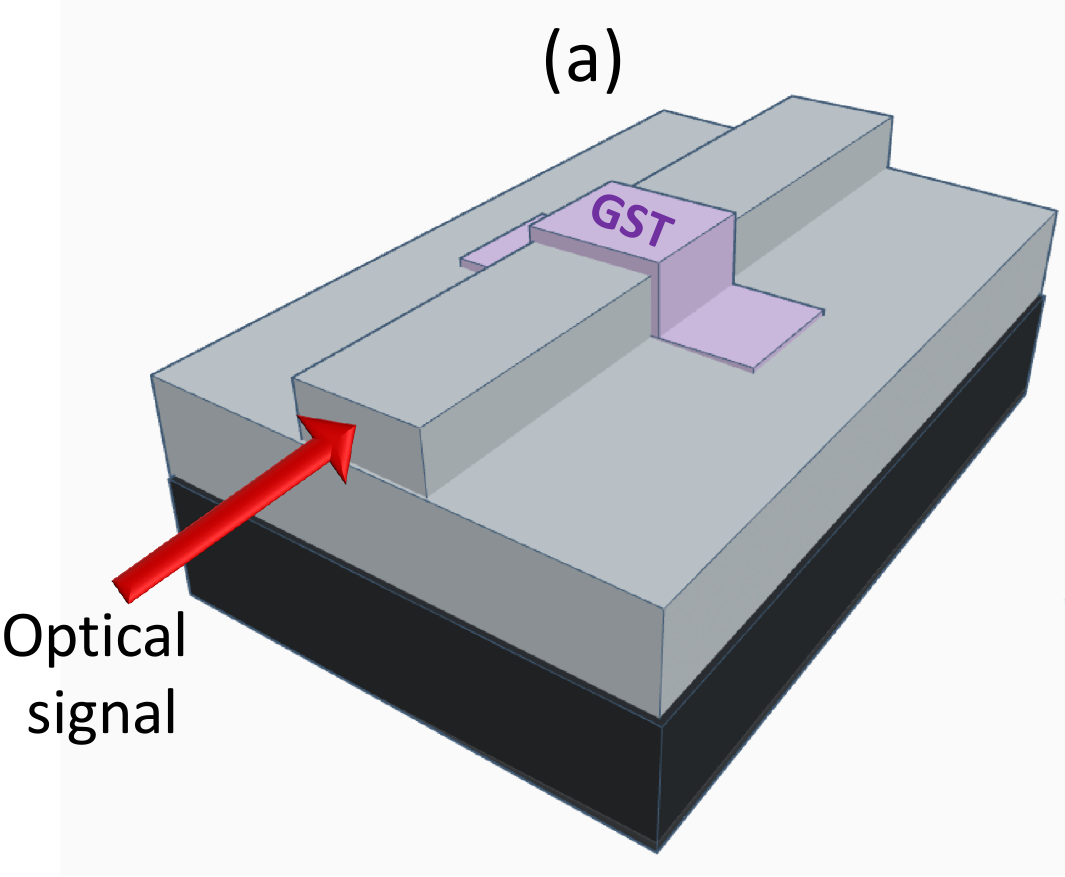}}
         \subfloat{
             \label{fig:OPCM_cross}
             \includegraphics[height=0.7in]{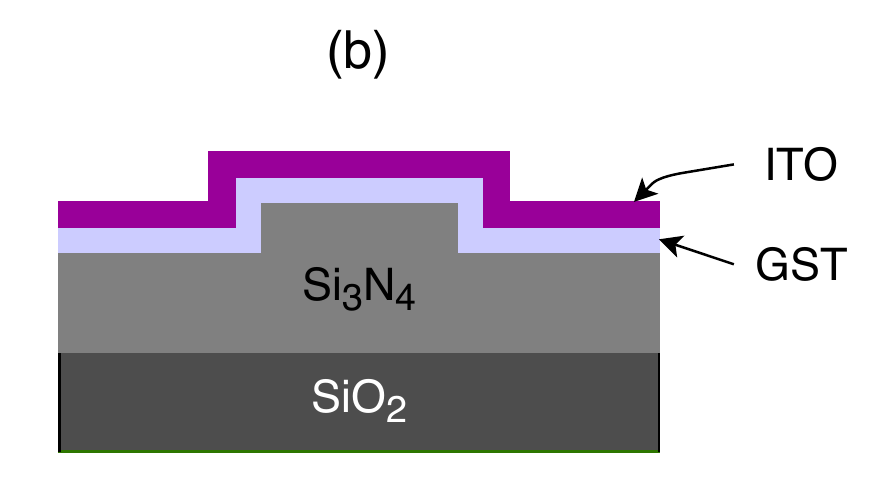}}
\caption{(a) 3D view of GST-based PCM cell. (b) Cross-sectional view of GST deposited on a $Si_3N_4$ waveguide.}
    \vspace{-0.2in}
\label{fig:pcm_integration}
\end{figure}

$Ge_2Sb_2Te_5$ (GST) is a well-known phase change material that exhibits high contrast in electrical property (resistance) and optical property (refractive index) between its two states, in addition to long data retention time and nanoscale size~\cite{wuttig2007phase, lyeo2006thermal, rios2014chip}. 
Thus, GST has been widely used as a storage element in a PCM cell (EPCM and \optmemory cells).
An \optmemory cell consists of only a GST element, and does not use a separate access transistor as an EPCM cell.
Figure~\ref{fig:pcm_integration} shows the structure of an \optmemory cell, where the GST is integrated on a waveguide~\cite{rios2015integrated, li2019fast}. 
The waveguides are fabricated using a $Si_3N_4$ layer deposited over a $SiO_2$ layer~\cite{li2020experimental}.
The GST layer is covered with a layer of Indium-Tin-Oxide (ITO) to prevent oxidation. 
The optical signals to read and write the \optmemory cell lie in the C band ($1530nm-1565nm$) and L band ($1565nm-1625nm$) of the telecommunication spectrum.

\subsection{Write Operation in \optmemory Cells}
For write operation, i.e., SET or RESET, the optical signal traversing through the waveguide is coupled to the GST element.
The energy of this optical signal heats the GST element and triggers a state transition.
For RESET operation, i.e., switching the GST element to an amorphous state (a-GST), an optical pulse of $180pJ$ energy is applied to the GST element for $25ns$~\cite{li2019fast}.
For SET operation, i.e., switching the GST element to a fully crystalline state (c-GST), an optical pulse with an energy of $130pJ$ is applied to the GST element for $250ns$~\cite{li2019fast}.
The transition of the GST state to a partially crystalline state requires different values of pulse energies ($60-130pJ$) applied for varying durations ($50-250ns$)~\cite{li2019fast}.

\subsection{Read Operation in \optmemory Cells}
The readout mechanism for an \optmemory cell uses the high contrast in the refractive indices of a-GST ($3.56$) and c-GST ($6.33$)~\cite{michel2014reversible}.
When an optical signal is passed through the GST element, the higher refractive index of c-GST results in an increased optical absorption by the GST element.
Rios {\it et al.}~\cite{rios2015integrated} demonstrate that c-GST absorbs $79\%$ of the input optical signal and allows transmission of only $21\%$ of the optical signal.
In contrast, a-GST transmits $100\%$ of the optical signal.
The transmission of partially crystalline states lies between $100\%$ and $21\%$~\cite{rios2015integrated}.
An \optmemory cell is, therefore, read out by sending a sub-$ns$ optical pulse through the GST element and measuring the transmitted optical intensity of the output pulse.
This transmitted intensity corresponds to a pre-determined bit pattern, thus allowing the readout of the stored data in the GST element.


\subsection{High MLC Capacity of \optmemory Cells}
In \optmemory cells, the read operation uses the refractive index of the GST state to determine the stored value.
Unlike the resistance value used in EPCM cells, the refractive index experiences minimal to no drift over time~\cite{li2019fast, rios2015integrated}.
This enables designing \optmemory cells with multiple stable partially crystalline states with each having unique refractive index.
Prior works have demonstrated that it is possible to reliably program an \optmemory cell to contain more than $34$ unique partially crystalline states~\cite{li2019fast, youngblood2019tunable}, which enables an \optmemory cell to have an MLC capacity of up to $5$ $bits/cell$.
Using a higher capacity MLC enables the read and write operation of a higher number of bits per access, thereby increasing the memory throughput.

\subsection{Silicon-Photonic Links}
In a computing system that uses a main memory composed of \optmemory cells, optical signals in silicon-photonic links can directly read/write the cells.
The silicon-photonic links provide higher bandwidth density at negligible data-dependent power compared to electrical links~\cite{barwicz2007silicon, krishnamoorthy2015energy, Batten2012Jetcas}.
In addition, these silicon-photonic links have single-cycle latency, in contrast to electrical links that often take $3-4$ cycles each for a memory request and a memory response.
Moreover, we can multiplex multiple optical signals (up to $32$ signals) in a single waveguide, resulting in dense WDM~\cite{lee2008ultrahigh}.
MicroRing Resonators (MRRs) can modulate these optical signals at data rates up to $12Gbps$~\cite{Alduino_sipho_wdm, pandey2018four, thonnart2020popstar} giving a peak memory throughput of $384Gbps$ per link.
Therefore, it is possible to design densely multiplexed silicon-photonic links that can directly access the \optmemory cells, further increasing the memory throughput.

\section{Motivation}
\label{sec:motivation}

\begin{figure}[t]
    \centering
    \includegraphics[width=0.9\columnwidth]{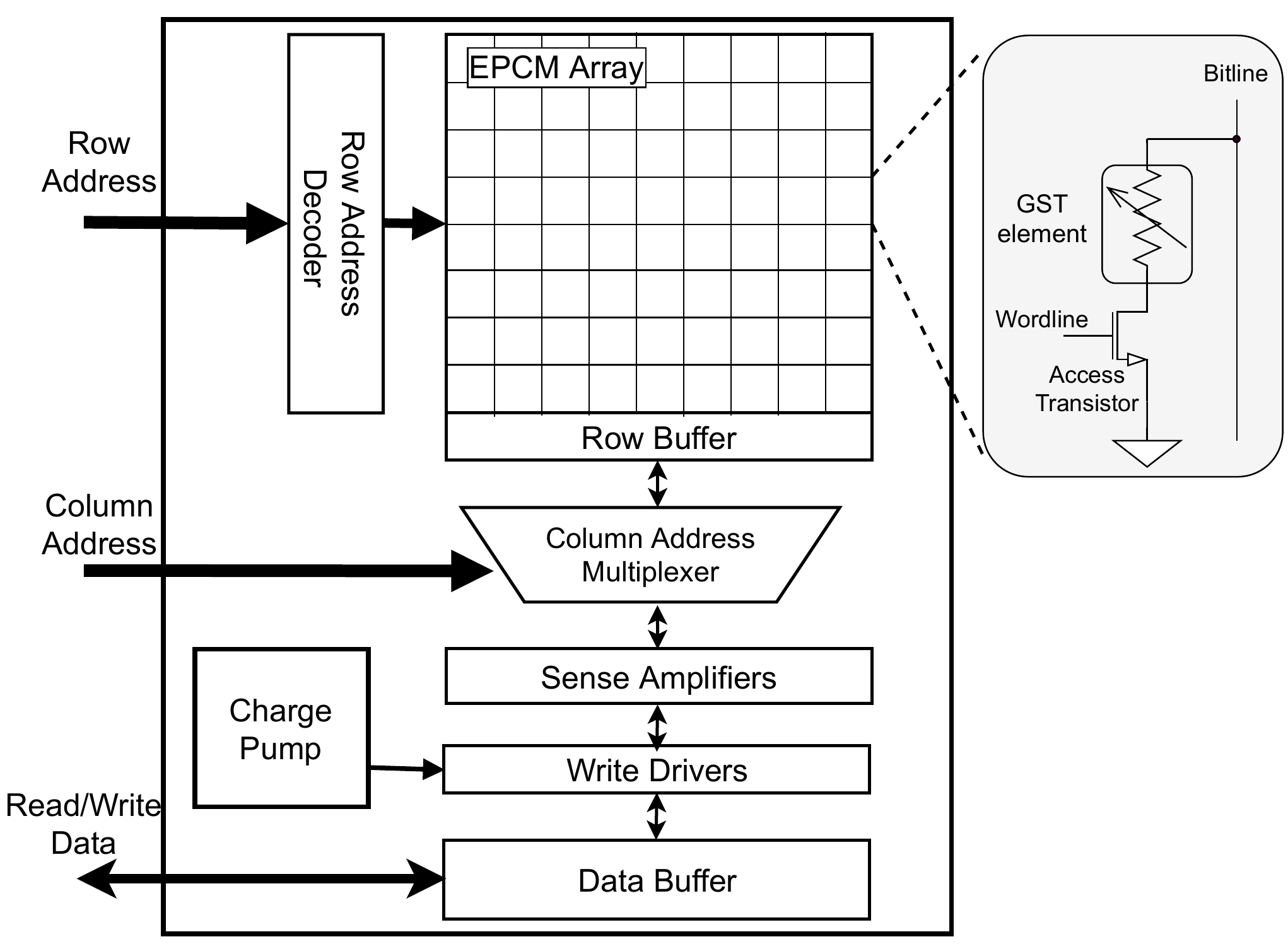}
    \caption{A typical EPCM architecture~\cite{lee2009architecting}.}
    \label{fig:epcm_array_arch}
    \vspace{-0.1in}
\end{figure}

\label{ssec:epcm_array}
In this section, we first describe the typical EPCM architecture and then explain why such an architectural design is impractical for \optmemory arrays.
Figure~\ref{fig:epcm_array_arch} shows the architecture of EPCM~\cite{lee2009architecting,kim2019ll}.
The EPCM array is a hierarchical organization of banks, blocks, and sub-blocks~\cite{lee2009architecting}.
During read or write operation, the EPCM first receives a row address.
The row address decoder reads the appropriate row from the EPCM array into a row buffer.
The EPCM next receives the column address, and the column address multiplexer selects the appropriate data block from the row buffer.
The bitlines of the selected data block are connected to the write drivers for write operation or to the sense amplifiers for read operation.
For write operation, the charge pumps supply the required drive voltage to the write drivers, which corresponds to SET or RESET operation.
For read operation, a read current is first passed through the GST element in the EPCM cell through an access transistor~\cite{lee2009architecting}.
Then, sense amplifiers determine the voltage on the bitline to read out logic $0$ or logic $1$.

Naively adapting the EPCM architecture for \mbox{\optmemory}, by just replacing the EPCM cells with \mbox{\optmemory} cells raises latency, energy and thermal concerns, thereby rendering such a design impractical. 
To understand these concerns, let us consider an \mbox{\optmemory} array that uses the EPCM architecture from Figure\mbox{~\ref{fig:epcm_array_arch}} with either an optical row buffer or an electrical row buffer.
Such an \mbox{\optmemory} array architecture has following limitations:

\textbf{Limitations with optical row buffer:}
An optical row buffer can be designed using a row of GST elements, whose states are controlled using optical signals.
When a row is read from the \mbox{\optmemory} array using an optical signal, the data is encoded in the signal's intensity.
This intensity is not large enough to update the state of the GST elements in the optical row buffer.
So the read value first needs to be converted into an electrical signal.
Based on this value, an optical signal with the appropriate intensity is generated to write the value into the optical row buffer.
%
Essentially we perform an extra O-E and E-O conversion.
This necessitates the use of photodetectors, receivers, transmitters and optical pulse generators, which adds to the energy and latency of a memory access.
Hence, an optical row buffer is not a viable option.

\indent \textbf{Limitations with electrical row buffer:}
An electrical row buffer can be designed either using capacitor cells as in DRAM or using phase change materials controlled using electrical current as in EPCM.
In both cases, the row buffer is accessed using electrical signals.
This increases the access latency and energy, and creates thermal issues as follows:
\begin{enumerate}[leftmargin=\parindent,align=left,labelwidth=\parindent,labelsep=0pt]
\item \textbf{Impact on read latency:} 
Upon receiving a row address from the MC on electrical links, the address first needs to be converted to an optical pulse, which is then used to read data from \optmemory cells.
After optical readout of an entire row from \optmemory array, the data has to be converted back into electrical domain to store it in the row buffer.
These two operations require an E-O and an O-E conversion, respectively, inside the \optmemory array.
These additional E-O/O-E conversions adds a latency of $25-30$ cycles for each read access\mbox{~\cite{bahadori2016energy}}.
\item \textbf{Impact on write latency:} 
When writing data from the row buffer to the \optmemory array, a set of sense amplifiers reads the data from the electrical row buffer.
This row buffer data is then mapped onto optical signals with appropriate intensities using a pulse generation circuitry within memory.
The optical signals are then used to write the data to the \optmemory cells.
Therefore, the write operation requires three E-O/O-E conversions, which adds a latency of $40-45$ cycles for each write access\mbox{~\cite{bahadori2016energy}}.
\item \textbf{Impact on read/write energy:} 
The energy spent in the peripheral circuitry for optical signal generation and readout, as well as in the circuitry for E-O-E conversion increases the active power dissipation within memory~\cite{notomi2014toward, bahadori2016energy,narayan2019waves}.
Since each read/write operation encounters multiple E-O-E conversions, the energy per read and write access rises considerably high ($>200pJ/bit$)~\mbox{\cite{dong2012nvsim}}. 
\item \textbf{Thermal issues:} 
The MRRs used in the \optmemory array are highly sensitive to thermal variations~\cite{padmaraju2014resolving}.
The thermal variations due to active electrical circuits within memory lowers the reliability of the MRR operation.
Such a design calls for active thermal and power management in \mbox{\optmemory}, which contributes to a power overhead of $10-30W$\mbox{~\cite{abellan2016adaptive}}.

\end{enumerate}

Furthermore, using silicon-photonic links in combination with \optmemory requires additional E-O and O-E conversions with this EPCM architecture that exacerbate the above discussed problems.
Hence, we argue for the need to redesign the microarchitecture and the read/write access mechanisms that are tailored to the properties of the \optmemory cell technology and the associated silicon-photonic link technology.
\section{\opcmsystem Architecture}
\label{sec:opcm_arch}

\begin{figure*}[t]
\centering
\includegraphics[width=\textwidth]{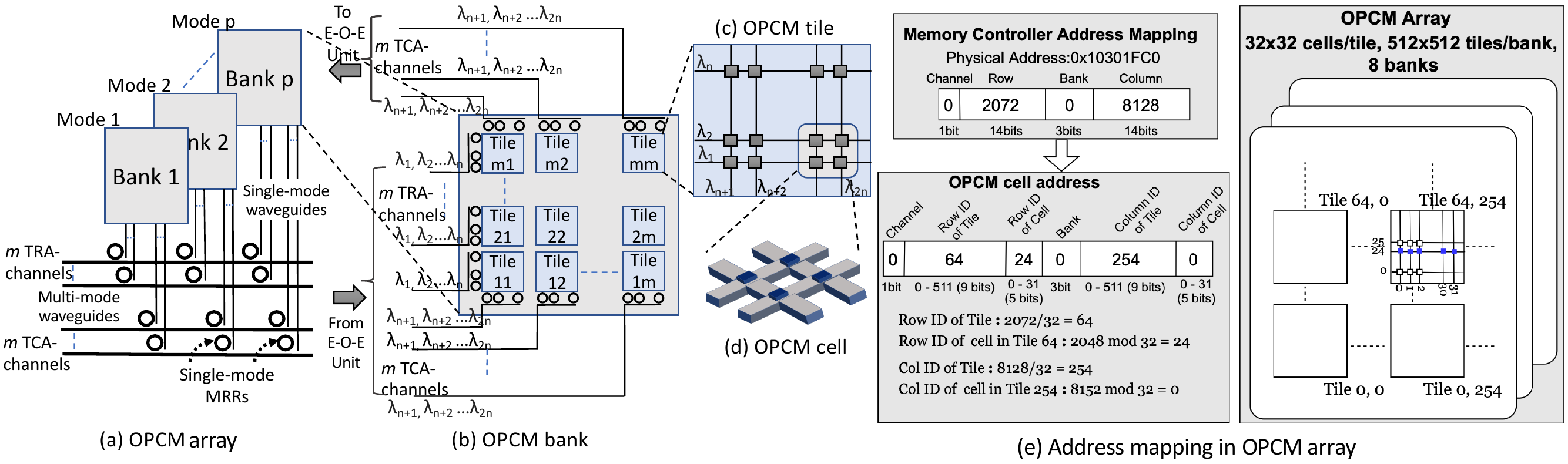}
\caption{(a) A multibanked-\optmemory uses $p$ optical modes to access $p$ banks.  (b) An \optmemory bank is an array of $m\times m$ tiles. Every tile is accessed by a TRA-channel and a TCA-channel, each channel containing $n$ optical signals.  (c) An \optmemory tile is an array of $n\times n$ cells. Every cell is accessed by a unique pair of optical signals. (d) \optmemory cells are placed at every waveguide crossing. (e) Address mapping of the physical address to cells in the \optmemory array. The physical address corresponds to \optmemory cells in the shaded blue row of \optmemory array.}
\vspace{-0.1in}
\label{fig:opcm_arch}
\end{figure*}

In this section, we describe the microarchitecture of the high-throughput \optmemory array in \opcmsystem.
\textbf{The key innovation of our proposed microarchitecture is enabling direct access of \optmemory cells by the optical signals in the silicon-photonic links.}
This direct access avoids the extra E-O and O-E conversions that are required if we were to adapt the EPCM architecture for \opcmsystem. 
Our \optmemory array microarchitecture is a hierarchical multi-banked design that maximizes the degree of parallelism for read and write accesses within the array using a combination of WDM and MDM.
A distinguishing feature of our {\optmemory} array design is that it does not contain any active circuits that consume power, i.e., it only contains passive optical devices.
Figure~\ref{fig:opcm_arch} illustrates the detailed microarchitecture of our proposed \optmemory array in \opcmsystem that uses GST as the phase change material. 
We describe each component of the proposed architecture, particularly focusing on how to access an {\optmemory} cell in the optical domain with minimal E-O and O-E conversions.


\subsection{\optmemory Tile}
\label{ssec:opcm_tile}
An \optmemory tile (see Figure~\ref{fig:opcm_arch}c) consists of an $n \times n$ array of GST elements, i.e., \optmemory cells. 
The GST elements are placed on top of waveguide crossings as shown in Figure~\ref{fig:opcm_arch}d.
This organization enables every {\optmemory} cell to be accessed using a unique pair of optical signals: one on the associated row and one on the associated column.
We need a total of $n$ unique optical signals with wavelengths $\lambda_1$, $\lambda_2$,..., $\lambda_n$ that are routed in the rows (one per row waveguide), and 
$n$ unique optical signals with wavelengths $\lambda_{n+1}$, $\lambda_{n+2}$,..., $\lambda_{2n}$ that are routed in the columns (one per column waveguide).
Wavelengths $\lambda_1$ to $\lambda_n$ together form the Tile Row Access (TRA)-channel, and wavelengths $\lambda_{n+1}$ to $\lambda_{2n}$ together form the Tile Column Access (TCA)-channel. 
A TRA-channel (and similarly each TCA-channel) is mapped to one or more waveguides depending on the number of wavelengths that can be multiplexed in a waveguide.
Owing to MLC, each \optmemory cell stores $b_{cell}$ bits. 
The total capacity of an \optmemory tile is $n^2.b_{cell}$.
A maximum of $n$ cells can be read/written in parallel from a single tile, which gives us a peak throughput of $n.b_{cell}$ bits per read/write access for a tile.

\subsection{\optmemory Bank}
\label{ssec:opcm_bank}
Figure~\ref{fig:opcm_arch}b shows the organization of an \optmemory bank. 
The \optmemory bank is composed of an array of $m\times m$ \optmemory tiles, and has a total capacity of $m^2.n^2.b_{cell}$ bits.
The \optmemory bank uses $m$ TRA-channels, one for each row in the bank, and $m$ TCA-channels, one for each column in the bank to communicate with the \eoe unit.
Each TRA-channel uses $\lambda_1$ to $\lambda_n$, and each TCA-channel uses $\lambda_{n+1}$ to $\lambda_{2n}$.
We design a hierarchical array of \optmemory cells ($m^2$ tiles with $n^2$ {\optmemory} cells per tile) instead of a large monolithic array ($m^2.n^2$ \optmemory cells), as designed by Feldman \mbox{\it et al.}~\mbox{\cite{feldmann2017calculating, feldmann2019integrated}} to decrease the laser power by the optical signals.
With our proposed design, the laser sources only need to support $2n$ unique optical signals (in the range of $\lambda_1$ to $\lambda_{2n}$) instead of the $m.2n$ unique optical signals that would be required in a large monolithic array.
We utilize MRRs to couple the optical signals of each TRA-channel and TCA-channel to its corresponding tile.
We need $n$ MRRs that are tuned to $\lambda_1$ to $\lambda_n$ in each of the $m$ TRA-channels and $n$ MRRs that are tuned to $\lambda_{n+1}$ to $\lambda_{2n}$ in each of the $m$ TCA-channels.

\subsection{Multi-banked \optmemory Array}
Figure~\ref{fig:opcm_arch}a shows the proposed multi-banked organization of the \optmemory array using MDM.
We interleave a cache-line across multiple banks.
There are $p$ banks, each supporting one of the $p$ spatial modes of the $2n$ optical signals. 
Bank $1$ only uses mode $1$ of all optical signals $\lambda_1$,.. $\lambda_n$ and $\lambda_{n+1}$,.. $\lambda_{2n}$, Bank $2$ only uses mode $2$ of all optical signals, and so on.
The waveguides connecting the \optmemory to the \eoe unit are multi-mode waveguides, which carry all the $p$ spatial modes of optical signals.
We employ single-mode MRRs~\cite{yang2014silicon, wang2017chip} that couple a single spatial mode of optical signals from the multi-mode waveguide to a bank.
Multiple prior works have exploited MDM property of optical signals coupled with WDM to design high-bandwidth-density silicon-photonic links~\cite{luo2014wdm, wu2017mode}.

\subsection{Address Mapping in \opcmsystem}
\label{ssec:address_mapping}
{\NEW Figure~\ref{fig:opcm_arch}e shows an example mapping of the physical address received by the MC to the physical location of cells within the \optmemory array in \opcmsystem.
A cache line of $64B$ is stored in a total of $128$ \optmemory cells with $4bits/cell$.
We interleave the cache line across $4$ different banks.
Within a bank, we map the 128-bit chunk of a cache line to a tile.
The tile has $32 \times 32 $ cells, and so we map that 128-bit chunk to an entire row within a tile.
The row (column) field of physical address in the MC is mapped to the row ID of tile (column ID of tile) field and the row ID of cell (column ID of cell) field.
In Figure~\ref{fig:opcm_arch}e, we show how the different fields of the physical address \texttt{0x10301FC0} are mapped to bank ID, row ID of tile, column ID of tile, row ID of cell, and column ID of cell.
}

\section{Access Protocol in \opcmsystem}
\label{sec:read_write}
To enable high-throughput access of \optmemory cells within the \optmemory array, we propose a novel read and write access protocol for \opcmsystem.
When the MC issues a read or write operation, the row address and column address are entered into the Row Address Queue and Column Address Queue, respectively, and the write data is entered into the Data Buffer in the \eoe unit.

\subsection{Writing a cache line to \optmemory array}
\label{ssec:write_opcm}

\ignore{\NEW To write a cache line to the \optmemory array in \opcmsystem, the \eoe unit identifies the bank ID, the row ID and column ID of the tile, and the row ID and column ID of \optmemory cell within a tile using our proposed address mapping.
In our example with $32 \times 32$ array of cells in a tile, a write operation updates all the cells in a row since a cache line is spread across a row within a tile (any misaligned accesses are handled on the processor side).
The \eoe unit determines the optical intensity that is required at each \optmemory cell in the row to write the cache line.
It then breaks down the optical intensity into two signals, one with a constant intensity of $I_0$ and the other with a data-dependent intensity of $I_i$, $i=1,2,..,128$.
The \eoe unit then modulates this constant optical intensity $I_0$ onto the optical signal corresponding to the row within a tile.
It then modulates the data-dependent optical intensities $I_1$, $I_2$, ...,$I_{128}$ onto the optical signals corresponding to the columns within that tile.
The superposition of the optical signals, i.e., $I_0+I_1$, $I_0+I_2$, ...., $I_0+I_{128}$ updates the state of the \optmemory cells in that row to the desired data.
Note that since a cache line is spread across 4 banks, the \eoe unit modulates data on optical signals to write to an \optmemory tile in each of these 4 banks.
Moreover, none of the optical signals individually carries sufficient intensity to trigger a state transition at any cell, so none of the other cells along the row or column are affected.}

{\NEW
To write a cache line to the \optmemory array, the \eoe unit identifies the bank ID, the row ID and column ID of the tile, and the row ID and column ID of the cell within a tile using the address mapping.
In our example with $32 \times 32$ array of cells in a tile, when writing 128-bit chunk of a cache line, we end up updating all the cells in a row (any misaligned accesses are handled on the processor side).
Hence, for writes at cache line granularity, the column ID within a tile is not used.
The \eoe unit determines the optical intensity that is required at each \optmemory cell in the row to write the 128-bit chunk of the cache line.
It then breaks down the optical intensity into two signals, one with a constant intensity of $I_0$ and the other with a data-dependent intensity of $I_i$, where $i=1,2,...,128$.
The \eoe unit modulates the constant intensity $I_0$ onto the optical signal corresponding to the row (selected by the row ID of cell) within a tile.
The \eoe unit then modulates the data-dependent optical intensities (i.e., $I_1$, $I_2$, ...,$I_{128}$) onto the optical signals corresponding to the columns within the tile. 
The \eoe unit transmits the row signal $I_0$, and the column optical signals $I_1, I_2, ...,I_{128}$ in parallel to write the cache line in the \optmemory array.
The superposition of the optical signals, i.e., $I_0+I_1$, $I_0+I_2$, ..., $I_0+I_{128}$ updates the state of the \optmemory cells.
Note that since a cache line is spread across 4 banks, the \eoe unit modulates data on optical signals to write to an \optmemory tile in each of these 4 banks.
None of the optical signals individually carries sufficient intensity to trigger a state transition at any cell, so none of the other cells along the row or column are affected.
}

\begin{figure*}[t]
\centering
\includegraphics[width=\textwidth]{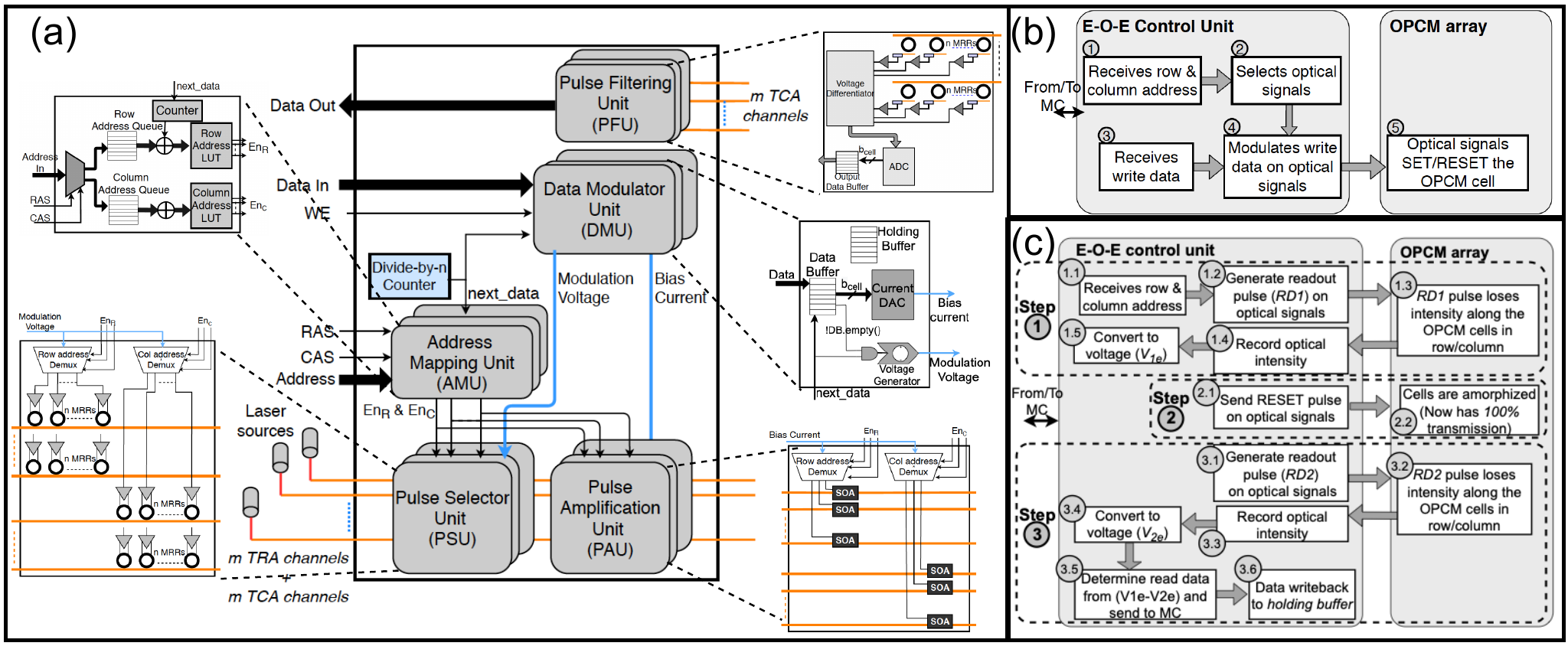}

\caption{\small (a) \eoe unit design. DMU: Generates the modulation voltage and the bias current corresponding to read/write data. AMU: Determines optical signals that correspond to read/write address. PSU: Selects the optical signals. PAU: Amplifies the optical signals using the bias current. PFU: Filters the optical signals to read cell data. Different micro-steps performed in \eoe unit and \optmemory array during (b) write operation and (c) read operation.}
    \vspace{-0.2in}

\label{fig:eoe_arch}
\end{figure*}

\subsection{Reading a cache line from \optmemory array}
\label{ssec:read_opcm}

{\NEW To read a cache line from \optmemory array, the \eoe unit transmits sub-ns optical pulses along all the columns in a tile that contain the cache line and measures the pulse attenuation.
However, there are multiple \optmemory cells along each column and so the output intensity of optical signals will be attenuated by all cells in that column.
It is, therefore, not possible to determine the \optmemory cell values using a one-pulse readout.
Hence, we use a three-step process for read operation of \optmemory array in \opcmsystem.
\begin{enumerate*}[label=\protect\circled{\arabic*}]
    \item To read a cache line, the \eoe unit first determines the bank ID, row ID and column ID of tile, row ID and column ID of cell.
    The \eoe unit transmits a read pulse $RD_1$ through all the columns in a tile containing the cache line.
    Note that since a cache line is spread across 4 banks, the \eoe unit transmits $RD_1$ on the $4$ different optical modes corresponding to the $4$ banks.
    Each read pulse is attenuated by all the \optmemory cells in the column.
    The attenuated pulses are received by the \eoe unit, which records the intensities of these attenuated pulses as $I_{1,1}$, $I_{2,1}$, ..., $I_{128,1}$. 
    These intensities are converted into electrical voltage and stored as $V_{1,1}$, $V_{2,1}$, ..., $V_{128,1}$.
    \item The \eoe unit then transmits a RESET pulse to the \optmemory cells of the cache line, i.e., all the cells along a row within a tile.
    All the cells along the row are now amorphized and have $100\%$ optical transmission.
    \item The \eoe unit then sends a second read pulse $RD_2$ through all the columns of a tile containing the cache line.
    Each read pulse is again attenuated by all \optmemory cells in the column.
    Given that step 2 amorphized all \optmemory cells of the cache line, the output pulse intensities are different from those in step 1.
    The attenuated pulses are received by the \eoe unit, which records the intensities of these attenuated pulses as $I_{1,2}$, $I_{2,2}$, ..., $I_{128,2}$. 
    These intensities are converted into electrical voltage and stored as $V_{1,2}$, $V_{2,2}$, ..., $V_{128,2}$.
    The \eoe unit computes the difference of the stored voltages of steps 1 and 3, i.e., $V_{1,1}-V_{1,2}$, $V_{2,1}-V_{2,2}$, ..., $V_{128,1}-V_{128,2}$.
    This difference is used to determine the cache line data stored in the \optmemory cells.
\end{enumerate*}}

\subsection{Opportunistic Writeback after Read}
\label{ssec:opportunistic_writeback}
{\NEW The RESET operation in step 2 of the read operation destructs the original data in the \optmemory cells.
We, therefore, perform an opportunistic writeback of the cache line to the \optmemory cells.
After completing the 3 steps of the read operation, the read data and the address are saved into a {\it holding} buffer in the \eoe unit.
When there are no pending read or write operations from the MC, the \mbox{\eoe} unit reads the data and its address from the {\it holding} buffer and writes the data back to the \mbox{\optmemory} array.
This writeback operation does not block any critical pending read and write operations coming from the MC.
The dependencies in read and write requests between the {\it holding buffer} and the data buffer is handled in the \eoe unit.
For a Read-After-Read case, the second read operation reads the data from the {\it holding buffer} if present.
If the data is not in the \textit{holding buffer} then the second read operation just uses the 3-step process + writeback (described above) to complete the read operation.
For a Write-After-Read case, if the write address matches the read address and there is an entry for that read in the {\it holding buffer}, then the corresponding entry in the {\it holding buffer} is invalidated. 
The write data is then  entered into the data buffer and then written into the appropriate \optmemory array.
The Write-After-Write and Read-After-Write are not an issue as the \eoe unit processes them in order.






}



\section{E-O-E Control Unit Design}
\label{sec:EOE_unit}

Our proposed \mbox{\eoe} unit provides the interface between the processor and the \optmemory array. 
The MC sends standard DRAM access protocol commands to the \mbox{\eoe} unit.
The \eoe unit maps these commands onto optical signals that read/write the data from/to {\optmemory} array. 
Given that {\optmemory} cells do not require Activate, Precharge and Refresh operations, the \mbox{\eoe} unit does not take any action for these commands.
Though we can design a {\opcmsystem}-specific MC and the associated read/write protocol, our goal is to enable the {\opcmsystem} operation with a standard MC in any processor.
The \eoe unit uses the following five sub-units to read from and write to the \optmemory array: data modulation unit (DMU), address mapping unit (AMU), pulse selector unit (PSU), pulse amplification unit (PAU), and pulse filtering unit (PFU).
Each \optmemory bank has a dedicated set of these five sub-units in the \eoe unit.
Figure~\ref{fig:eoe_arch}a shows the design of the \eoe unit in \opcmsystem and the internals of these sub-units.

Figure~\ref{fig:eoe_arch}b illustrates the sequence of operations in the \eoe unit for write operation to a bank containing $512 \times 512$ tiles with $32 \times 32$ cells per tile (same design as that used in Figure~\ref{fig:opcm_arch}e).
The AMU in the \eoe unit first receives the row address and then the column address from MC (Step 1).
Depending on the addresses, the PSU in the \eoe unit selects the appropriate optical signals using the address mapping explained in Section~\ref{ssec:address_mapping} (Step 2).
The PSU selects one optical signal and $32$ optical signals for writing to $32$ cells in a tile.
In parallel with the write address, the DMU in the \eoe unit receives the write data from the MC (Step 3).
The DMU generates a unique bias current for each of the $32$ optical signals depending on write data and applies the currents to the semiconductor optical amplifiers (SOA) in the PAU (Step 4).
The SOAs amplify the optical signals to the required intensities.
These amplified signals and the optical signal traverse through the silicon-photonic links to the appropriate \optmemory cells in the bank, and SET/RESET the cell (Step 5).
The \eoe unit incurs a latency of $T_{EO}$ cycles to map the address and data onto optical signals, resulting in a peak throughput of $1/T_{EO}$.

Figure~\ref{fig:eoe_arch}c illustrates the sequence of operations in the \eoe unit for the 3-step read operation from a bank.
In the first step, the AMU receives the row and column addresses from MC and selects the appropriate $32$ optical signals in the PSU using the address mapping explained in Section~\ref{ssec:address_mapping} (Step 1.1).
The DMU generates a low-intensity readout pulse ($RD_1$) and the PAU modulates this pulse on the $32$ optical signals (Step 1.2).
The optical signals traverse through the silicon-photonic link and then through the columns in the tile.
The optical signals lose intensity as they pass through all the \optmemory cells in their associated columns (Step 1.3).
The intensities of these attenuated signals are recorded by the PFU (Step 1.4).
The PFU then converts the optical intensities into electrical voltages, $V_{1,1}$, $V_{2,1}$, ..., $V_{32,1}$ (Step 1.5).
In the second step, the DMU generates an optical intensity that corresponds to a RESET pulse.
This RESET pulse is mapped onto the appropriate optical signals and these signals are sent to the \optmemory array (Step 2.1).
The signals traverse through the silicon-photonic links and amorphize the \optmemory cells corresponding to the read address (Step 2.2).
In the third step, the DMU generates another readout pulse ($RD_2$) and the PAU modulates this pulse on a set of $32$ optical signals (Step 3.1).
These signals traverse through the silicon-photonic links and then through the appropriate columns in the tile.
These signals too loses intensity as they pass through all the \optmemory cells in their associated columns (Step 3.2).
The PFU records these attenuated signals (Step 3.3) and converts these optical signals into electrical voltages $V_{1,2}$, $V_{2,2}$, ..., $V_{32,2}$ (Step 3.4).
Finally, the PFU computes $V_{1,1}-V_{1,2}$, $V_{2,1}-V_{2,2}$, ..., $V_{32,1}-V_{32,2}$ to determine the data (Step 3.5) and sends the data to the MC.
The PFU also writes this data back to the {\it holding buffer} in the DMU (Step 3.6).

\section{Evaluation Methodology}
\label{sec:eval_meth}

\subsection{Multicore System with \opcmsystem}
\label{ssec:comp_system}
We use an 8-core processor for our evaluation.
We primarily evaluate \opcmsystem with 4-bit MLC \optmemory cells (given that \optmemory cell with $5$ $bits/cell$ has been prototyped~\cite{li2019fast}) against an EPCM with $2$ $bits/cell$.
Table~\ref{tab:system_details} details the processor and memory configurations.
For processor-memory networks, we consider electrical as well as silicon-photonic links, with $1GT/s$ transfer rate per link. We obtain a peak bandwidth of $64GB/s$ in EPCM and $256GB/s$ in \mbox{\opcmsystem}.

The \optmemory array in \opcmsystem is organized as a single rank connected to a memory channel on the MC via the \eoe unit. 
Each one of the 8 \optmemory banks has its dedicated set of DMU, ATU, PSU, PAU, and PFU sub-unit in the \eoe unit.
The average SET latency is $t_{SET}$ + $t_{EOE}$, $165ns$, the RESET latency is $t_{RESET}$ + $t_{EOE}$, $30ns$, and the read latency is $t_{read}$ (time for 3-step read operation) + $t_{EOE}$, i.e., $30ns$.
A maximum of $t_{SET}/t_{EOE}=32$ writes can be issued from the \eoe unit to \optmemory in parallel. 
So, we can write $32\times b_{cell}$ bits in parallel.
A maximum of $t_{read}/t_{EOE}=5$ reads is issued from the \eoe unit to \optmemory in parallel.
So, we can read $5\times b_{cell}$ bits in parallel.
{\NEW We use a holding buffer that is large enough ($16$ cache line slots from our evaluations) to avoid stalling any read/write memory requests from the MC.}

\begin{table}[t]
\small
\caption{Architectural Details of the Simulated System.}
\label{tab:system_details}
\begin{tabular}{|P{1.5cm}|P{6.1cm}|}
\hline
\multicolumn{2}{|c|}{\textbf{Processor, On-chip caches}}          \\ \hline
Cores     & 8-core, 1GHz x86 ISA, Out-of-Order, 192 ROB entries, dispatch/fetch/issue/commit width=8         \\ \hline
L1 caches &  32kB split L1 I\$ and D\$, 2-way, 2-cycle hit, 64B, LRU, write-through, MSHR: 4 instruction \& 32 data \\ \hline
L2 cache  &  Shared L2\$, 2MB, 8-way, 20-cycle hit, 64B, LRU, write-back,  MSHR: 32  (I \& D)              \\ \hline
\multicolumn{2}{|c|}{\textbf{Main memory ($2GB$)}}                        \\ \hline
EPCM~\cite{choi201220nm}      &   4 banks, 8 devices/rank, 1 rank/channel, \newline 
              bus width = $64$, burst length = $4$           \newline
              $t_{SET}=120ns$, $t_{RESET}=50ns$,  $t_{read}=60ns$, $t_{BURST}=4ns$
\\ \hline
\optmemory array in \opcmsystem~\cite{rios2015integrated, li2019fast}     &   8 banks, 1 rank/channel, 1 device/rank  \newline 
              bus width = $32\times b_{cell}$, burst length = $8$           \newline
               $t_{SET}=160ns$, $t_{RESET}=25ns$,  $t_{read}=25ns$, $t_{BURST}=1ns$, $t_{EOE}=5ns$
\\ \hline
\end{tabular}
\end{table}


\subsection{Simulation Framework}
\label{ssec:sim_frame}
We model the architectural specifications of the system in gem5~\cite{binkert2011gem5}.
We conduct full-system simulations in gem5 with Ubuntu 12.04 OS and Linux kernel v4.8.13.
We fast-forward to the end of Linux boot and execute each workload for $10$ billion instructions.
The main memory models for DDR4 are based on DRAMSim2~\cite{rosenfeld2011dramsim2}.
For modeling EPCM and \optmemory, we integrate NVMain2.0~\cite{poremba2015nvmain} in gem5. 
\vspace{-0.06in}

\subsection{Workloads}
\label{ssec:workloads}
We simulate graph applications from GAP-BS benchmark \cite{beamer2015gap} and HPC applications from NAS-PB benchmark~\cite{bailey1991parallel}. 
We evaluate the graph applications on three different input datasets from SNAP repository~\cite{snapnets}: Google web graph ($google$), road network graph of Pennsylvania ($roadNetPA$), and Youtube online social network ($youtube$).
For HPC applications from NAS-PB benchmark, we use the large dataset. 
We execute 8 threads of these applications in a workload.
\section{Evaluation Results}
\label{sec:revision_results}

\subsection{\opcmsystem vs EPCM}
\label{ssec:epcm_perf}

\begin{figure}[b]
\centering
\includegraphics[width=\columnwidth]{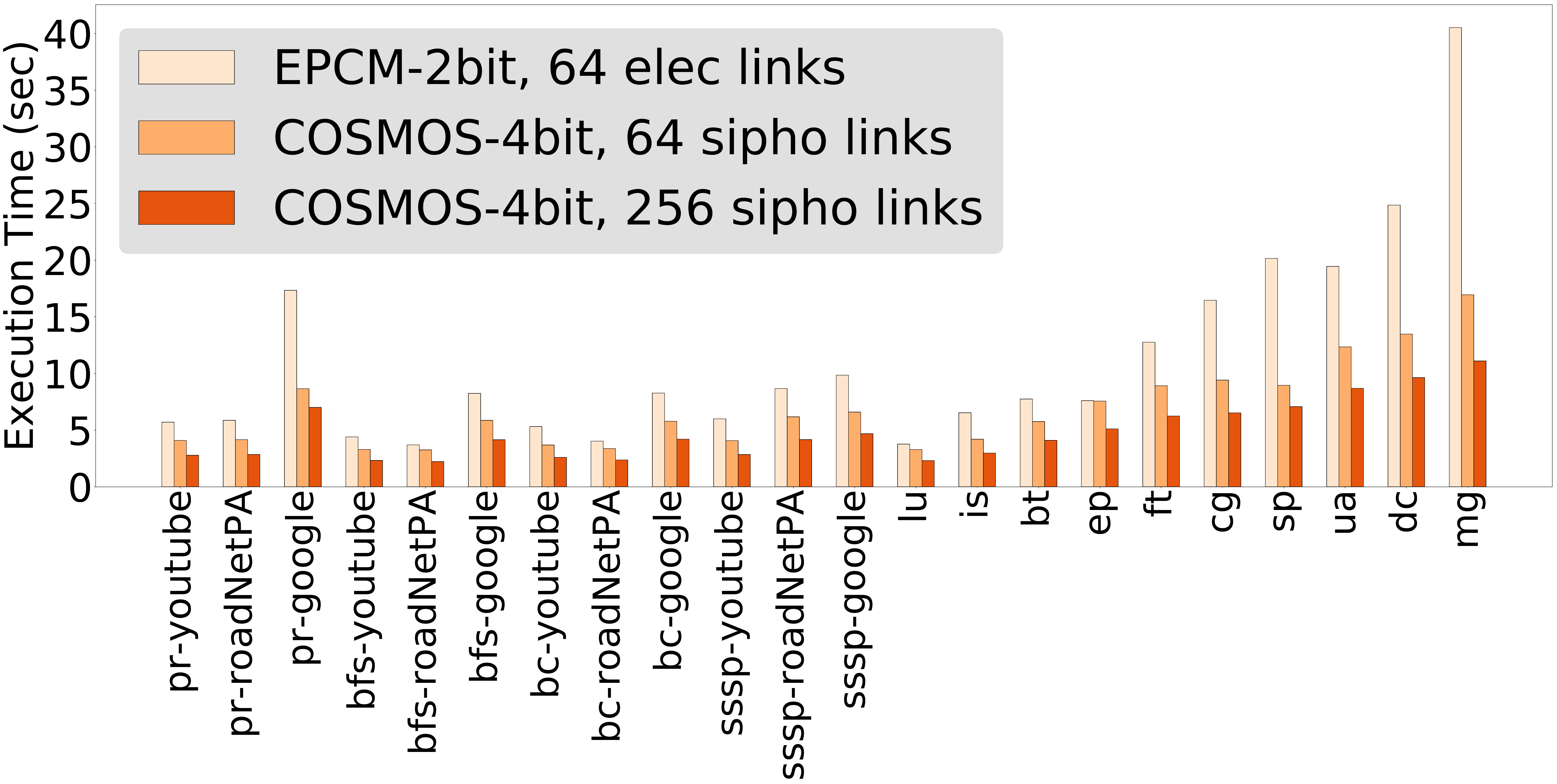}
\caption{  Performance comparison of \opcmsystem with EPCM.  }
\label{fig:epcm_opcm_perf}
\end{figure}

\begin{figure*}[t]
    \centering
        \subfloat{
            \includegraphics[height=1.2in]{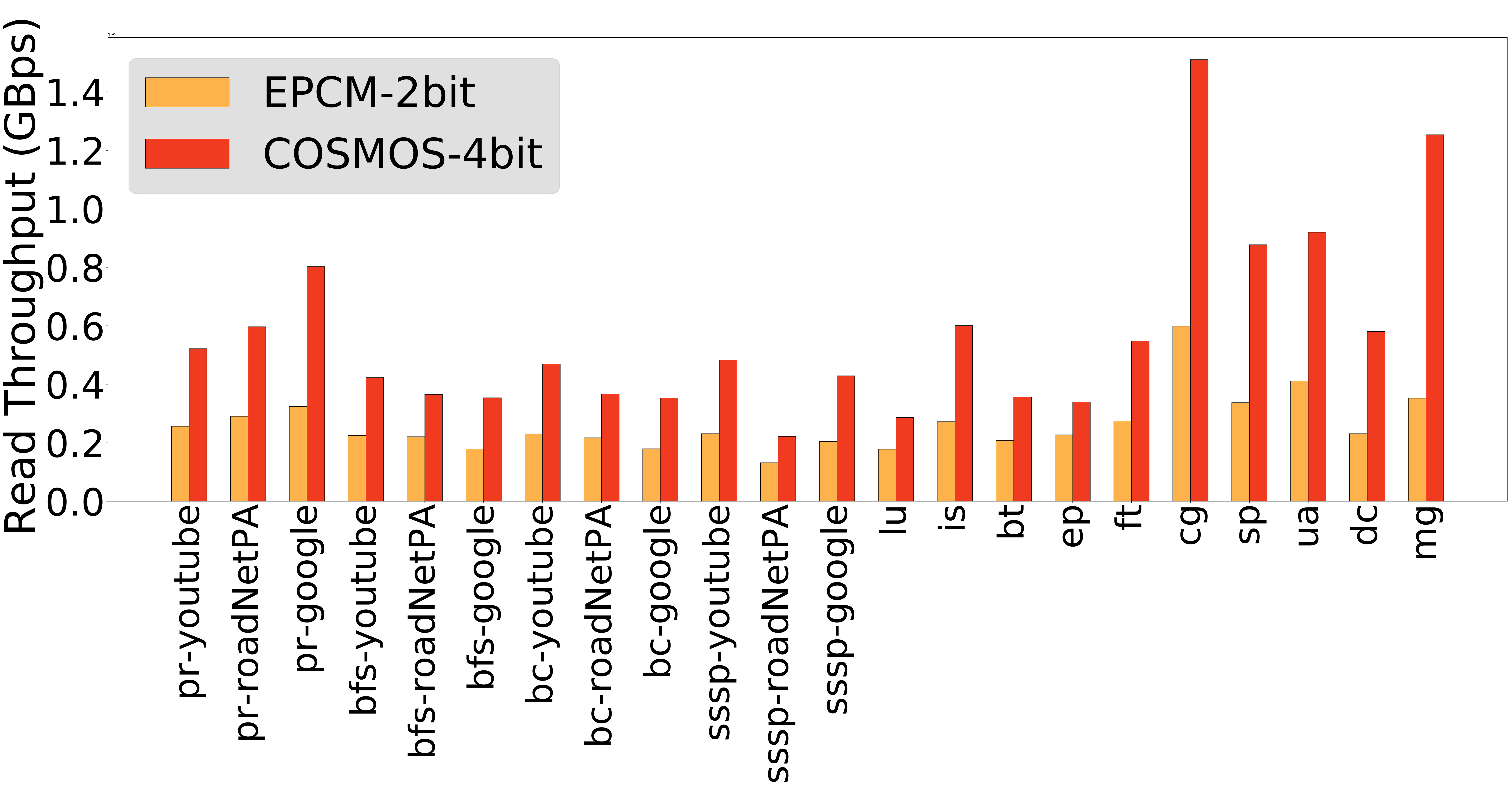}}
        \subfloat{
             \includegraphics[height=1.2in]{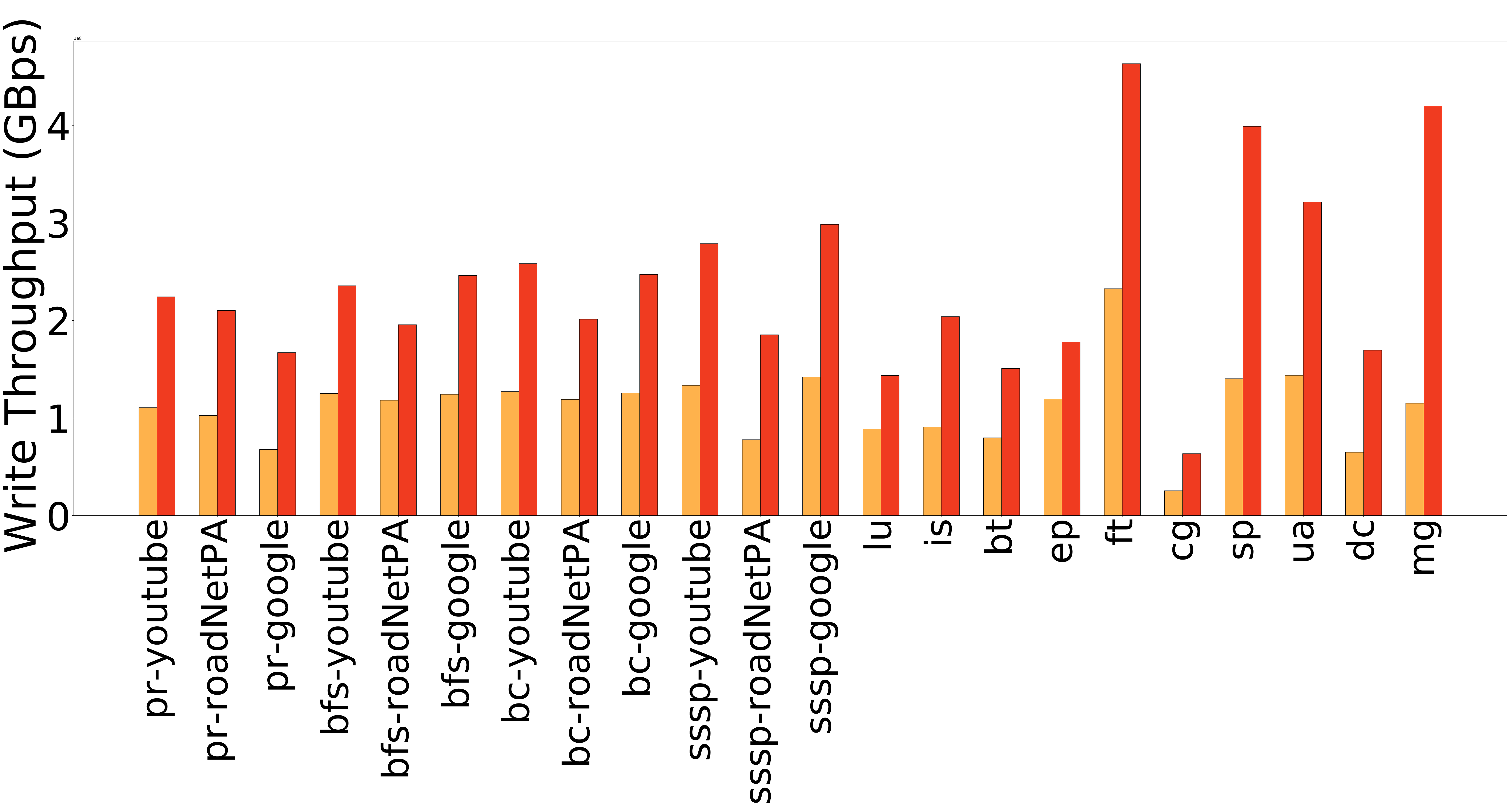}}
        \subfloat{
             \includegraphics[height=1.16in]{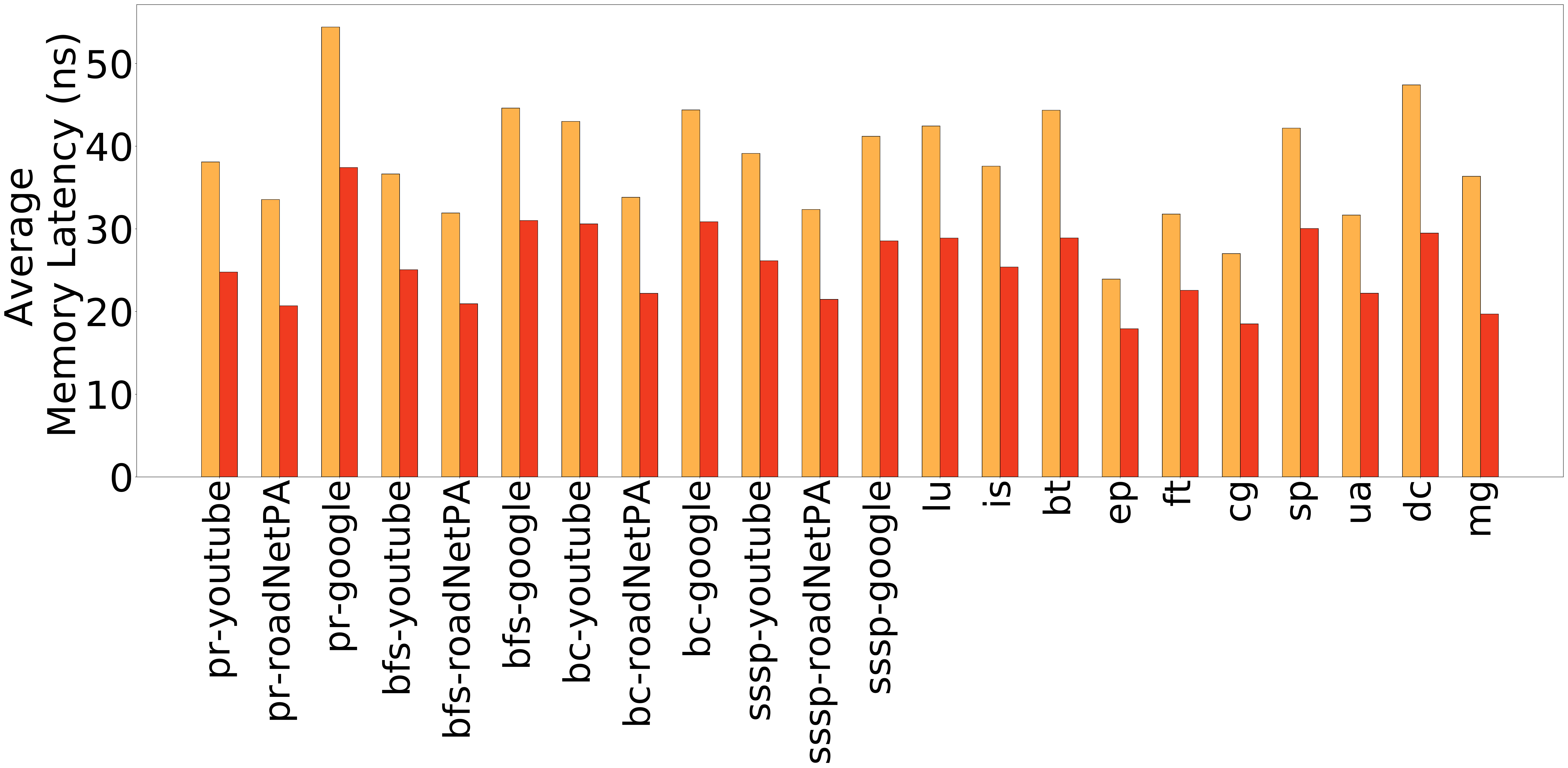}}
\caption{(a) Read throughput, (b) Write throughput, (c) Average memory latency}
\vspace{-0.15in}
\label{fig:thpt_latency}
\end{figure*}

\subsubsection{Performance}
We compare EPCM (2bit MLC or EPCM-2bit) that uses $64$ proces\-sor-to-memory electrical links with \opcmsystem (4bit \optmemory cells, or \opcmsystem-4bit) that also uses $64$ processor-to-memory silicon-photonic links, and with \opcmsystem-4bit that uses $256$ processor-to-memory silicon-photonic links.
Figure~\ref{fig:epcm_opcm_perf} shows the overall performance (execution time in seconds) for systems with these three configurations. 
Compared to the EPCM-2bit with 64 electrical links, \opcmsystem-4bit with 64 silicon-photonic links has on average $1.52\times$ better performance across all workloads.
This performance improvement is due to the higher $bits/access$ throughput of \opcmsystem resulting from higher MLC capacity and the single-cycle latency in silicon-photonic links.
Increasing the number of silicon-photonic links from $64$ to $256$ further improves the system performance.
Compared to EPCM-2bit using $64$ electrical links, we observe performance improvement of $2.14\times$ on average for graph and HPC workloads with \opcmsystem-4bit using $256$ silicon-photonic links.
These performance benefits are due to denser WDM in silicon-photonic links. 
\textit{The key takeaway from this comparison is that even though the \optmemory cells suffer from long write latency similar to  EPCM cells, the superior MLC capacity of \optmemory cells that are directly accessed by high-bandwidth-density silicon-photonic links improves the system performance in \opcmsystem.}

\subsubsection{Throughput}
Figures~\ref{fig:thpt_latency}a and \ref{fig:thpt_latency}b show the read and write throughput, respectively, of \opcmsystem-4bit with $256$ silicon-photonic links, and EPCM-2bit with $64$ electrical links.
Compared to EPCM-2bit with $64$ electrical links, \opcmsystem-4bit with $256$ silicon-photonic links theoretically has $8\times$ higher peak throughput, i.e., $2\times$ due to higher MLC capacity and the $4\times$ due to the increased number of processor-to-memory links.
Therefore, it is possible to issue increased number of parallel read and write operations in \opcmsystem-4bit.
From figure~\ref{fig:thpt_latency}a and figure~\ref{fig:thpt_latency}b we observe that \opcmsystem-4bit has $2.09\times$ higher read throughput and $2.15\times$ higher write throughput, respectively, than EPCM-2bit for graph and HPC workloads.
This increased read and write throughput of \opcmsystem-4bit hides the long write latencies.
Figure~\ref{fig:thpt_latency}c shows that the average memory latency (read+write) of \opcmsystem-4bit is $33\%$ lower than EPCM-2bit across all workloads.
\textit{The key insight from this study is the increased read and write throughput provided by the higher MLC capacity and the silicon-photonic links hides the long write latencies of \optmemory cells in \opcmsystem.}

\begin{table}[t]
\small
  \centering
  \caption{Optical power budget for $2GB$ \opcmsystem. The table shows optical power losses and SOA gain along the optical path from laser source to \optmemory cells.}
  \label{tab:optical_budget}
  \begin{tabular}{|P{3.8cm}|P{1.95cm}|P{1.4cm}|}
    \hline
    \textbf{Loss/gain component} & \textbf{Single} & \textbf{Total} \\
    \hline
    \hline
    Coupling loss & $-1dB$ & $-1dB$~\cite{batten2009building}  \\
    \hline
    MRR drop loss (\eoe) & $-0.5dB$~\cite{grani2014design} & $-0.5dB$ \\
    \hline
    MRR through loss (\eoe) & $-0.05dB$~\cite{grani2014design} & $-3.2dB$ \\
    \hline
    Propagation loss (Laser to SOA) & $-0.3dB/cm$~\cite{shang2015low} & $-0.09dB$ \\
    \hline
    SOA gain & $+20dB$ & $+20dB$ \\
    \hline
    Propagation loss (SOA to \optmemory) & $-0.3dB/cm$~\cite{shang2015low} & $-0.09dB$ \\
    \hline
    Bending loss & $-0.167dB$~\cite{shang2015low} & $-0.167dB$ \\
    \hline
    MRR drop loss (\optmemory) & $-0.5dB$~\cite{grani2014design}  & $-0.5dB$ \\
    \hline
    MRR through loss (\optmemory) & $-0.05dB$~\cite{grani2014design} & $-3.2dB$ \\
    \hline
    Propagation loss (in \optmemory) & $-0.03dB/cm$~\cite{shang2015low} & $-4.91dB$ \\
    \hline
    Max. power required to SET the GST & $\frac{135pJ}{250ns}$~\cite{li2019fast} & $-2.67dBm$ \\
    \hline
    \hline
    Power per optical signal & &$-7.22dBm = 0.19mW$ \\
    \hline
    Laser wall-plug efficiency & &$20\%$ \\
    \hline
    \hline
    \textbf{Total laser power} & &\textbf{$16.38W$} \\
    \hline
  \end{tabular}
  \vspace{-0.1in}
\end{table}

\subsubsection{Energy Consumption}
\label{ssec:energy_est}
{\NEW
The primary contributors to the overall power consumption during the read and write operations are the different active components in the \eoe unit and the laser sources that drive the silicon-photonic links.
The \optmemory array in \opcmsystem consists of only passive optical devices, so it does not consume any active or idle power.
The electrical power consumed in the laser source is proportional to its optical output power, which in turn depends on the optical losses in the path of the optical signal and the minimum power required to switch the farthest GST element.
Table~\mbox{\ref{tab:optical_budget}} lists the optical losses in the various components and the maximum switching power required at the GST element in decibels (dB).
The various optical losses and SOA gains are obtained from prior characterization works~\cite{batten2009building,grani2014design,shang2015low,li2019fast}.
By accounting for the wall-plug efficiency, we calculate the minimum required laser power per optical signal as $0.95mW$.
Aggregating the laser power for all optical signals required in a $2GB$ \opcmsystem system, we get a total laser power of $16.38W$.\footnote{We assume all laser sources are ON all the time. Development of laser power management techniques is left for future work.}
}

\begin{table}[t]
\small
\caption{Energy-per-bit for read and write accesses.}
\label{tab:energy-per-bit}
\centering
\begin{tabular}{|P{3.2cm}|P{1.6cm}|P{2cm}|}

\hline
      \textbf{Energy-per-bit (pJ/bit)}   & \textbf{EPCM-2bit} & \textbf{\opcmsystem-4bit}      \\ \hline \hline
\textbf{Write}  & 243   & 40.68   \\ \hline
\textbf{Read}   & 44.5  & 11.6 \\ \hline
\textbf{Opportunistic Writeback}   & NA  & 40.68 \\ \hline
\end{tabular}
\end{table}

{\NEW
In the \eoe unit, the current-DAC in DMU and the ADC in PFU consume $0.3mW$ each~\cite{rekhi2019analog}.
For \optmemory-4bit, $32$ write operations can be issued in parallel per bank, i.e., we can write $32\times b_{cell}\times 8=128B$ in parallel with an average write latency of $160ns$. 
That aggregates to writing 2 cache lines of $64B$ each in parallel.
A cache line is interleaved across 4 banks and is row aligned in an \optmemory tile.
Therefore, we need $4$ row optical signals and $4\times 32$ column optical signals to write a cache line.
Therefore, the total power of the laser, SOAs and DACs in the \eoe unit for writing 2 cache lines in parallel aggregates to $334.8mW$. This equates to $40.68pJ/bit$ for writing to \opcmsystem-4bit.

For read operation, up to 5 read operations can be issued in parallel per bank, i.e., $5\times b_{cell}\times 8=20B$ bits in parallel, with a read latency of $25ns$.
The total power of the laser, SOA, DAC, and ADC in \eoe for 5 parallel read operations is $9.3mW$, resulting in a read energy of $11.6pJ/bit$ for \opcmsystem-4bit.
The energy consumed in the electrical links connecting the processor and the \eoe unit is $<1pJ/bit$~\cite{coskun2020cross}.
For EPCM, we use NVSim~\cite{dong2012nvsim} to compute the energy-per-bit for read and write operations.
The opportunistic writeback operation in \opcmsystem uses the same energy as that required for write operation.
Table~\ref{tab:energy-per-bit} shows the energy-per-bit for EPCM-2bit and \opcmsystem-4bit.
The read and write energy-per-bit of \opcmsystem-4bit are $3.8\times$ and $5.97\times$ lower, respectively, than that of EPCM-2bit.

}

\subsection{Sensitivity Analysis of \opcmsystem}
\label{ssec:sensitivity_analysis}

\subsubsection{MLC values}
\label{sssec:mlc_result}

\begin{figure}[t]
\centering
\includegraphics[width=\columnwidth]{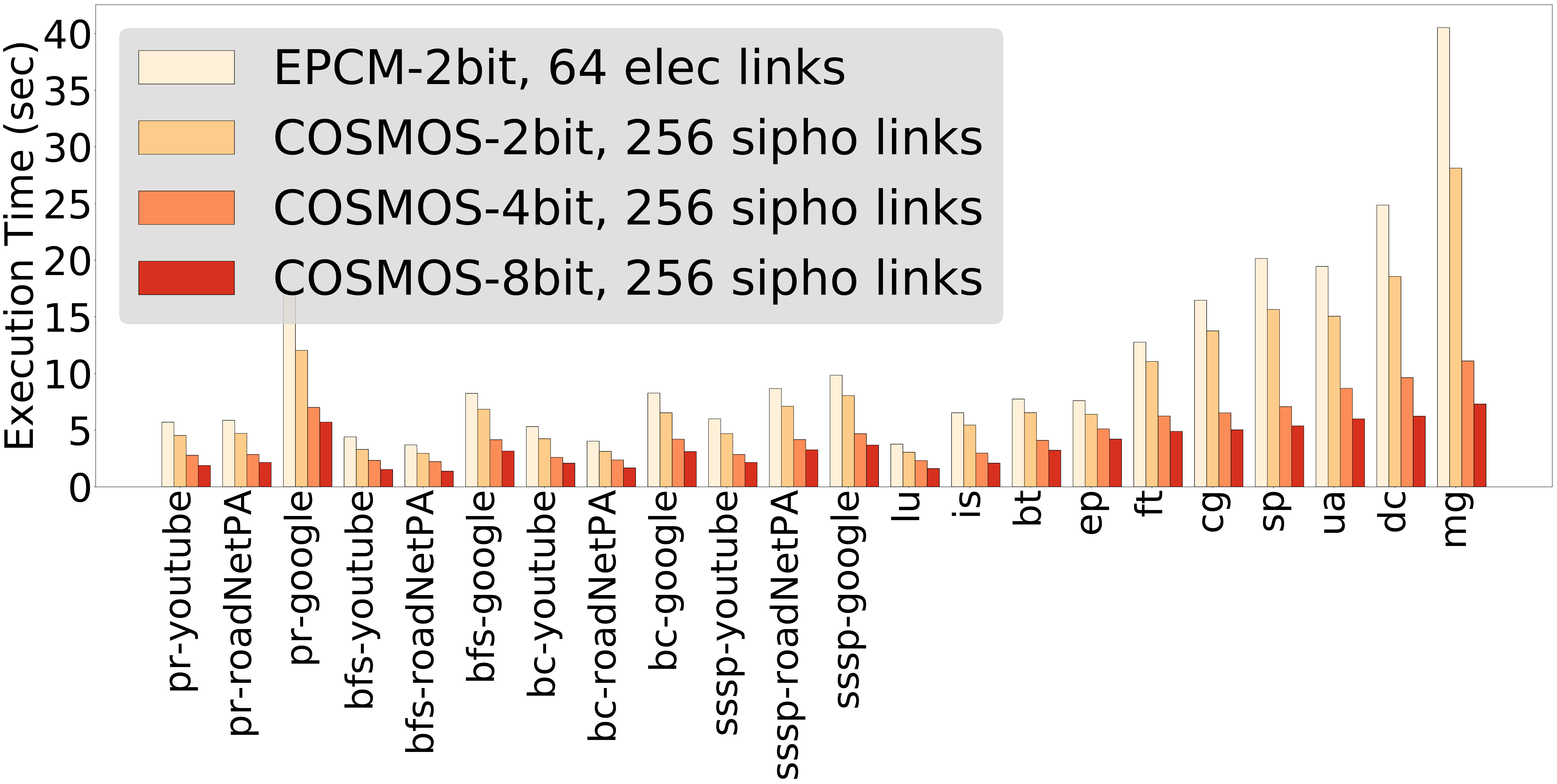}
\caption{Performance comparison of \opcmsystem with different MLC.  }
\vspace{-0.2in}
\label{fig:mlc-exectime}
\end{figure}

Rios \mbox{\it et al.} gave the first demonstration of a 2-bit {\optmemory} cell operation~\mbox{\cite{rios2015integrated}}. 
Advances in optical signaling and control have resulted in the demonstration of denser multilevel {\optmemory} cells. 
Li \mbox{\it et al.} demonstrated 5-6 bits per {\optmemory} cell~\mbox{\cite{li2019fast}}. 
Further prototypes have demonstrated scalable integration of {\optmemory} cell arrays in silicon and silicon nitride platforms~\mbox{\cite{li2020experimental, feldmann2019integrated}}.
With the maturity in optical integration technologies, \optmemory technology with $8$ $bits/cell$ is expected in the near future~\cite{li2019fast}.
We compare the systems performance of \opcmsystem that uses \optmemory cells with different MLC capacities, ranging from $2$ $bits/cell$ to $8$ $bits/cell$, for the same number of silicon-photonic links (see Figure~{\ref{fig:mlc-exectime}}).
The performance across applications increases, on average, by $39.2\%$ and $26.4\%$ as the MLC capacity of \optmemory cells increases from $2$ $bits/cell$ to $4$ $bits/cell$ and from $4$ $bits/cell$ to $8$ $bits/cell$, respectively.
An \optmemory cell with higher MLC capacity will provide higher memory throughput.

\subsubsection{Number of Silicon-Photonic Links}
\label{sssec:sipho_links_eval}

\begin{figure}[t]
\centering
\includegraphics[width=\columnwidth]{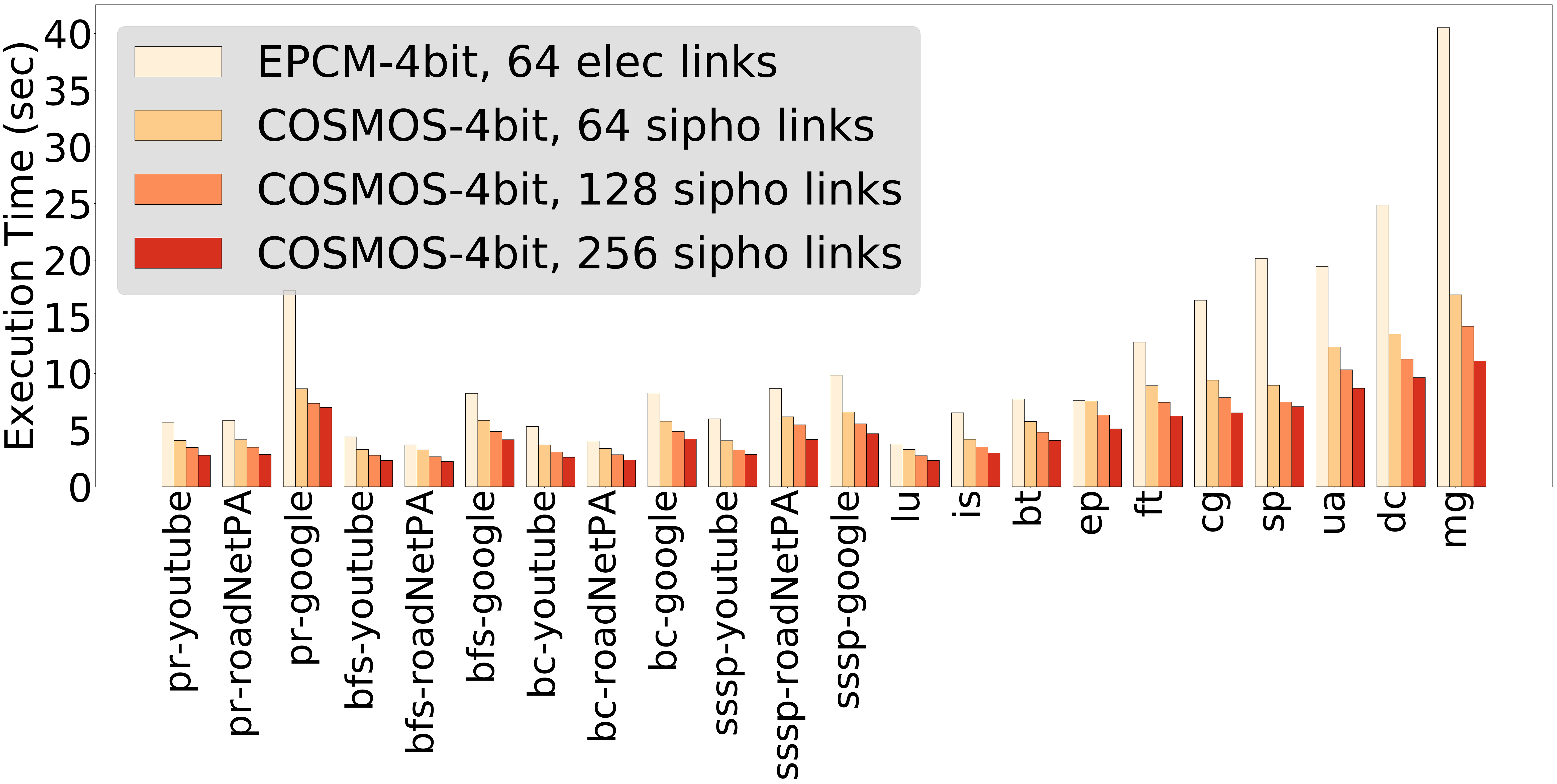}
\vspace{-0.25in}
\caption{ Performance comparison of \opcmsystem with different number of silicon-photonic links.}.  
\vspace{-0.2in}
\label{fig:sipho-exectime}
\end{figure}

We compare the performance of \opcmsystem-4bit with different number of silicon-photonic links (see Figure~\ref{fig:sipho-exectime}).
Multiplexing a higher number of optical signals in silicon-photonic links enables parallel read and write accesses of a higher number of \optmemory cells.
Due to this increased throughput, the overall system performance improves as the number of silicon-photonic links increases.
We observe a performance improvement of $29.3\%$ (on average) for \opcmsystem-4bit with $256$ silicon-photonic links over \opcmsystem-4bit with $64$ links.

\subsubsection{Holding Buffer}
\label{sssec:holding_buffer}
\begin{figure}[t]
\centering
\includegraphics[width=\columnwidth]{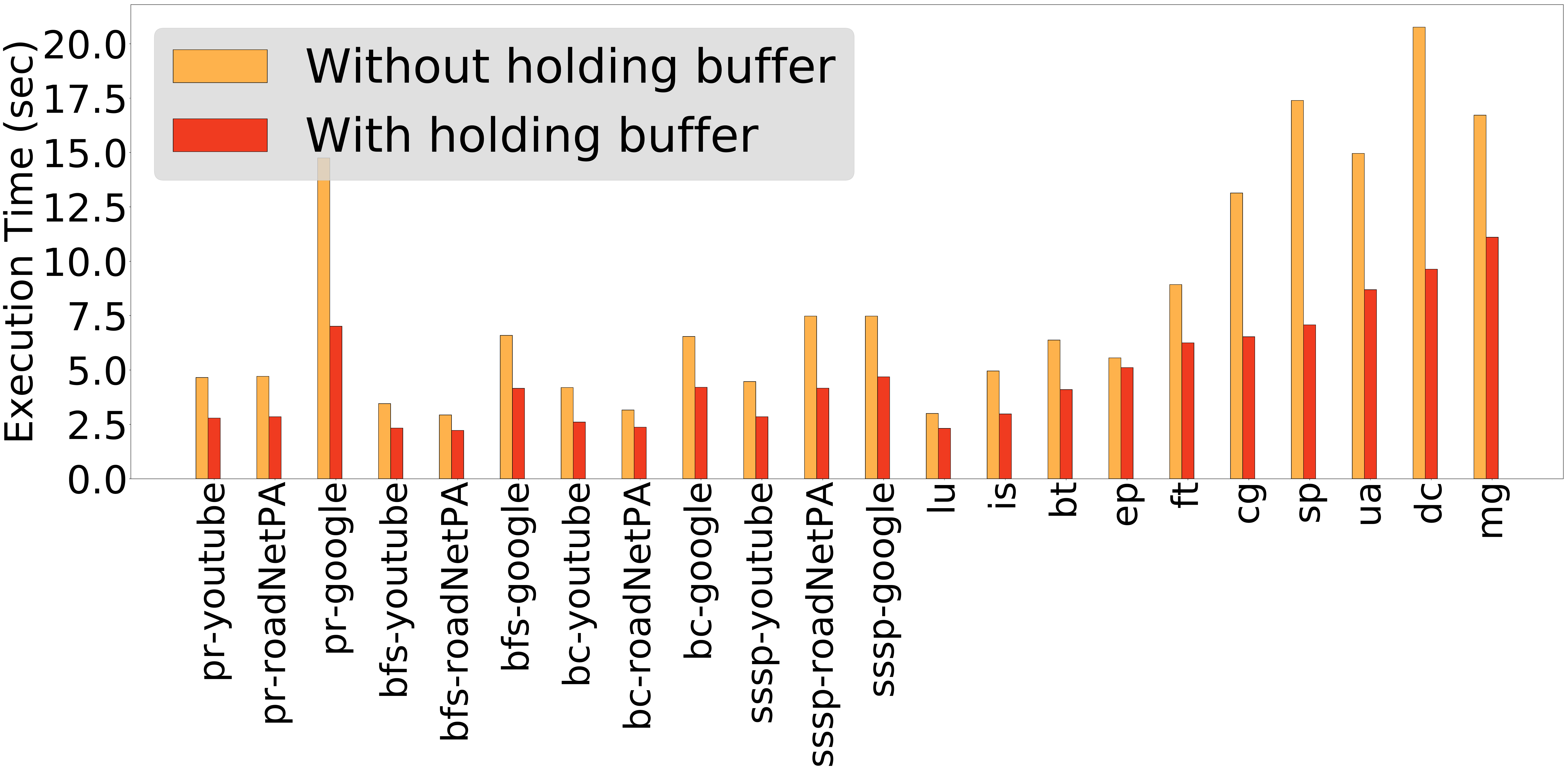}
\caption{Performance comparison of \opcmsystem with and without holding buffer for opportunistic writeback in read operation.  }
\label{fig:holding_buffer_eval}
\vspace{-0.2in}
\end{figure}

{\NEW 
Figure~\ref{fig:holding_buffer_eval} shows the system performance comparison with and without the {\it holding buffer}.
In absence of the {\it holding buffer}, the read data needs to be written back to the \optmemory cells immediately after readout because the read operation is destructive.
Therefore, the complete read operation incurs a total latency of readout latency ($25ns$) + writeback latency ($160ns$).
In contrast, when the \eoe unit consists of a {\it holding buffer}, the read data is stored in the {\it holding buffer} at the end of read operation.
The data from the {\it holding buffer} is written back to the \optmemory cells only when the DB in the \eoe unit is empty, ensuring that the writeback operation does not stall any critical read and write operations.
Using the highest read and write rate of the workloads that we evaluated, we determine that a {\it holding buffer} with $16$ cache line slots, i.e., $1KB$, is enough to avoid any memory read/write stalls.
The {\it holding buffer} occupies < 1000 $\mu m ^2$ area and can be integrated into the \eoe unit with minimal overhead.}

\subsection{Endurance Analysis of \opcmsystem}
\label{ssec:endurance}

Similar to EPCM, \optmemory cells have lower endurance due to cell wearout.
The \optmemory cell endurance depends on how often we write to that cell~\cite{qureshi2009enhancing}.
Given that the read operation in \opcmsystem also includes a write (RESET) in step 2, the endurance of \optmemory cells also depend on the read rate.
We determine the average write rate 
We estimate the \mbox{\opcmsystem} lifetime using the equation proposed by Qureshi \mbox{\it et al.}~\mbox{\cite{qureshi2009scalable}}:
\scalebox{1}{$Y=\frac{S.W_m}{B.F.2^{25}}$}
where, $Y$ is lifetime in years, $W_m$ is maximum allowable writes per cell ($10^6$ for OPCM cells\mbox{~\cite{rios2015integrated,li2019fast}}), $B$ is write rate in Bytes/cycle (average read+write rate across graph and HPC workloads), $F$ is core frequency in Hz ($1GHz$), and $S$ is \mbox{\opcmsystem} size in bytes ($2GB$, $4GB$ and $8GB$).

\begin{figure}[t]
    \centering
    \includegraphics[width=0.75\columnwidth]{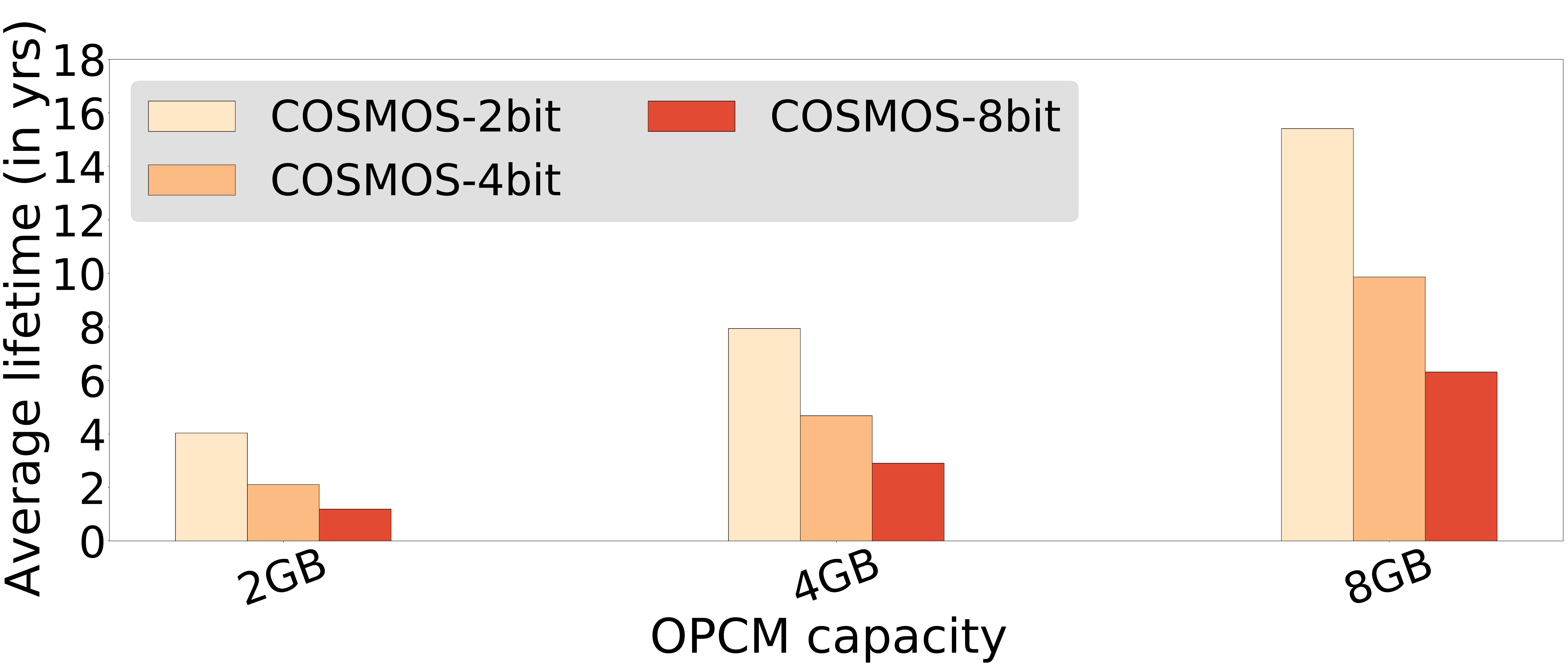}
    \vspace{-0.1in}
    \caption{{Average lifetime (in years) of \opcmsystem with different MLC capacities of \optmemory cells and different memory capacities.}}
    \vspace{-0.1in}
    \label{fig:endurance}
\end{figure}

Figure~{\ref{fig:endurance}} plots the average lifetime for {\optmemory} with different MLC capacities. 
Here, we assume that for a given memory size, all MLC options use the same number of silicon-photonic links. 
Hence, the \opcmsystem with 8-bit \optmemory cells has higher effective throughput than the \opcmsystem with 4-bit \optmemory cells and so an application running on \opcmsystem-8bit runs faster than an application running on \opcmsystem-4bit. 
As a result, for an application, even if the absolute number of memory writes is same for both \opcmsystem-8bit and \opcmsystem-4bit, the number of $writes/second$ to \opcmsystem-8bit is higher than the number of $writes/second$ to \opcmsystem-4bit.
Hence, the lifetime of \opcmsystem-8bit is lower than that of the \opcmsystem-4bit and \opcmsystem-2bit.

\begin{table}[t]
\small
\centering
 \caption{Dimensions of optical devices in the \mbox{\optmemory} array.}
 \label{tab:parameters}
\begin{tabular}{|P{4cm}|P{3cm}|}
\hline
\textbf{Optical device} &\textbf{Dimension}                         \\ \hline
GST &      $500nm \times 500nm$~\cite{rios2014chip,rios2015integrated}                       \\ \hline
Separation between adjacent GSTs &    $50nm$~\cite{hosseini2014optoelectronic} \\ \hline
MRR diameter &   $5\mu m$~\cite{6487665}          \\ \hline
\end{tabular}
\end{table}

\begin{table}[t]
\small

 \caption{Bit density ($bits/mm^2$) of memory technologies.}
 \label{tab:area}
\begin{tabular}{|P{1.8cm}|P{3.5cm}|P{2cm}|}
\hline
\textbf{Memory technology} &\textbf{Area of $2GB$ memory} & \textbf{Bit density ($bits/mm^2$)}                          \\ \hline
\textbf{DDR4} &      $224mm^2$~\cite{ddr4area}         & $9.14MB/mm^2$                       \\ \hline
\textbf{HBM2.0} &    $91.99mm^2$~\cite{kim2014hbm}           & $22.26MB/mm^2$ \\ \hline
\textbf{EPCM-2bit} &   $336mm^2$ (simulated~\cite{dong2012nvsim})  &      $6.095MB/mm^2$           \\ \hline
\textbf{3D OPCM-4bit} & $268.43mm^2$ (calculated) & $7.63MB/mm^2$ \\ \hline
\textbf{3D OPCM-8bit} & $67.1mm^2$ (calculated) & $30.52MB/mm^2$    \\ \hline
\end{tabular}
\end{table}

\subsection{Area Analysis of the \optmemory Array}
\label{ssec:area_eval}
To design the \mbox{\optmemory} array in \mbox{\opcmsystem}, we use the prototype of a GST element developed by \mbox{Rios {\it et al.}~\cite{rios2014chip,rios2015integrated}} and the MRR dimensions from prior work as shown in \mbox{Table~\ref{tab:parameters}}.
We use 3D stacking for \mbox{\optmemory} array, with different banks stacked vertically (one bank per layer).
The multi-mode waveguides in the interposer are routed vertically, and at each layer single-mode MRRs filter out the mode of all optical signals that belong to its corresponding bank.
For a $2GB$ 4-bit \mbox{\optmemory} array with $8$ banks, a single bank consists of $1024$ tiles with $32$ cells/tile and a row and column of MRRs as shown in \mbox{Figure~\ref{fig:opcm_arch}b}.\footnote{The tile size is limited by the number of unique optical signals in C and L bands with sufficient guardbands ($32$ in our case).
The number of banks depends on the number of unique electromagnetic modes that can be supported ($8$ in our case).}
A bank, therefore, is composed of $1024\times32$ GSTs along a row/column with $(1024\times32-1)\times 50nm$ of separation between GSTs, and a single row/column of MRRs at the beginning.
Using the dimensions of these optical devices listed in \mbox{Table~\ref{tab:parameters}}, we calculate the area of a $2GB$ \mbox{\optmemory} array and its bit density and report it in \mbox{Table~\ref{tab:area}}.

We compare the area and bit density of the 3D-stacked \optmemory array in \opcmsystem with DDR4, 3D-stacked HBM2.0 and EPCM-2bit memory system (see Table~\ref{tab:area}).
With current \optmemory cell footprints, 3D-stacked {\optmemory}-4bit has $1.2\times$ and $2.9\times$ lower bit density than DDR4 and HBM2.0, respectively, and $1.25\times$ higher bit density than EPCM-2bit.
3D-stacked {\optmemory}-8bit has $3.4\times$, $1.4\times$ and $5\times$ higher bit density than DDR4, HBM2.0 and EPCM-2bit, respectively.
Nevertheless, device-level research efforts have demonstrated that GST elements are highly scalable and can retain the electrical and optical characteristics at amorphous and crystalline states~\cite{5388621,9133093}.
An aggressive chip prototype with $200nm \times 200nm$ GST element with $50nm$ separation has been recently fabricated~\cite{hosseini2014optoelectronic}.
These aggressive optical fabrication technologies promise achieving several orders higher densities for \optmemory arrays than current DRAM technologies.

\begin{table*}[t]
\small
\caption{Survey of research efforts to improve write performance and write energy for using EPCM as main memory. The performance gains and energy reductions are shown in comparison to a naive EPCM system. (NR: Not reported)}
\vspace{-0.1in}

  \label{tab:relatedefforts}
\begin{tabular}{|P{1.4cm}|P{1.3cm}|P{1.4cm}|P{1.4cm}|P{1.3cm}|P{1.1cm}|P{1.3cm}|P{1.2cm}|P{1.2cm}|P{2.1cm}|}
\hline
  & Fine-grained power budgeting~\cite{jiang2012fpb} & Write truncation \cite{jiang2012improving} & Dynamic write consolidation \cite{xia2014dwc} & Logical decoupling \& mapping \cite{yoon2014efficient} & Proactive SET~\cite{qureshi2012preset} & Partition-aware scheduling \cite{song2019enabling} & Double-XOR mapping \cite{yu_bitmapping} & Boosting rank parallelism \cite{arjomand2016boosting} & \textbf{Optical control with silicon-photonic links}\\ \hline \hline
Performance gains            & $76\%$          & $26\%$         & $17.9\%$            & $19.2\%$        & $34\%$      & $28\%$   & $12\%$ & $16.7\%$ & \textbf{$2.31\times$} \\ \hline
Energy reductions          &    NR            &     NR          & $13.9\%$           & $14.4\%$       & $25\%$       & $20\%$   &   NR    &     NR    &   \textbf{$4\times$}     \\ \hline
\end{tabular}
\vspace{-0.20in}
\end{table*}

\subsection{\opcmsystem vs DRAM}
\label{ssec:dram_vs_opcm}

\begin{figure}[t]
\centering
\includegraphics[width=\columnwidth]{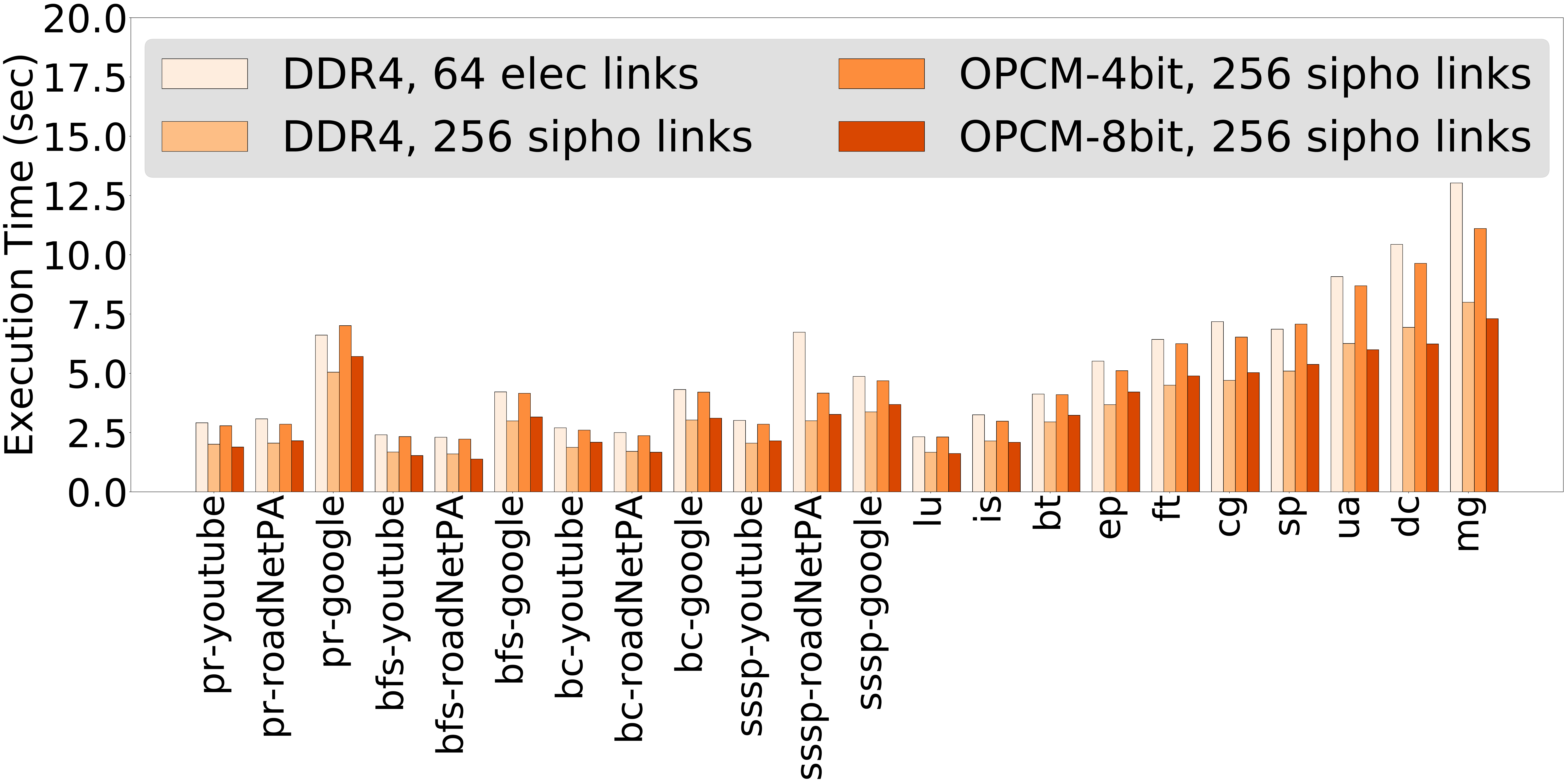}

\caption{  Performance comparison of \mbox{\optmemory} with DDR4 DRAM. } 
\label{fig:ddr4_opcm_perf}

\end{figure}

The overarching goal of \opcmsystem is to replace DRAM systems that are used widely in computing systems.
We noted that though all other NVM systems (in their current form) provide non-volatility, data persistence and high scalability, their poor performance negates their benefits and makes them impractical to replace DRAM systems.
We, therefore, compare the performance and energy of DDR4 with 64 electrical links, DDR4 with 256 silicon-photonic links~\cite{beamer2009re}, \opcmsystem-4bit with 256 silicon-photonic links, and \opcmsystem-8bit with 256 silicon-photonic links.
Figure~\ref{fig:ddr4_opcm_perf} shows the overall system performance across the four configurations.
For DDR4, replacing $64$ electrical links with $256$ silicon-photonic links provides $32\%$ average performance improvement.
This improvement results from the higher throughput due to dense WDM and single-cycle latency of silicon-photonic links.
With \opcmsystem-4bit, we obtain $5.6\%$ improvement in performance compared to DDR4 with $64$ electrical links.
This is in stark contrast to EPCM-2bit, which performs $3-4\times$ worse than DDR4.
\opcmsystem-8bit with $256$ silicon-photonic links performs $30.6\%$ better than DDR4 with $64$ electrical links and $2.1\%$ better than DDR4 with $256$ silicon-photonic links.
The increased read and write throughput due to the higher MLC capacity and dense WDM silicon-photonic links reduces the average memory access latency of \opcmsystem.
Figure~\ref{fig:thpt_latency}c shows the the average memory latency in \opcmsystem is $33.64ns$ across all workloads, which is lower than DDR4 DRAM ($40ns$).
Moreover, from Table~\ref{tab:energy-per-bit} we observe that energy-per-access for write operation in \opcmsystem-4bit is similar to that of DDR4 DRAM ($40pJ/bit$~\cite{ddr4density}) and the energy-per-access for read operation in \opcmsystem-4bit is $3.45\times$ lower than DDR4 DRAM ($40pJ/bit$~\cite{ddr4density}).

Though we evaluate DDR4 memory with silicon-photonic links, such a system encounters several design challenges.
To support silicon-photonic links in DDR4, memory requests from MC require an E-O conversion in MC and an O-E conversion in memory, and memory responses from DDR4 require an E-O conversion in memory and an O-E conversion in MC.
Effectively, we need two extra conversions on the memory side.
The active peripheral circuitry to support E-O-E conversions within memory increases the power density and raises thermal concerns.
Due to the high thermal sensitivity of MRRs, there is a need for active thermal management.
The power and resulting thermal concerns affect the reliability of optical communication in DRAM systems.

We observe that \opcmsystem with $4$ $bits/cell$ \optmemory array demonstrates similar performance and energy characteristics as current DDR4 systems, while \opcmsystem with $8$ $bits/cell$ \optmemory array improves performance.
This is particularly exciting as \opcmsystem exhibits zero leakage power, better scaling and non-volatility, making it a viable replacement for DRAM in the near future.

\section{Related Work}
\label{sec:related_work}

\subsection{Phase Change Memories}
Several works have proposed architectural and management policies to address the PCM challenges and have designed EPCM systems either as a standalone main memory, as part of hybrid DRAM-PCM systems or as a storage memory between DRAM and flash memory~\cite{qureshi2009scalable, lee2011energy, lee2011characterizing, jia2016dynamic, ramos2011page, jiang2012fpb, jiang2012improving, xia2014dwc, yoon2014efficient, qureshi2012preset,thakkar2017dyphase,kim2019ll,yu_bitmapping, arjomand2016boosting, song2019enabling}.
Most of these efforts have focused on addressing the long write latency and high write energy. A summary of these efforts is shown in Table~\ref{tab:relatedefforts}.
Hybrid DRAM-PCM systems leverage the higher bit density in PCMs for improved performance, but at the cost of higher write energy~\cite{qureshi2009scalable, lee2011energy, lee2011characterizing, jia2016dynamic, ramos2011page}.
To address PCM cell wearout, the techniques to enhance the write endurance include rotation-based wear leveling~\cite{qureshi2009enhancing}, process variation-aware leveling~\cite{dong2011wear, zhao2014slc}, and writeback minimization and endurance management~\cite{ferreira2010increasing}.
Due to lower write endurance, PCM cells are also susceptible to malicious write attacks.
Common strategies employed in EPCMs to thwart these attacks include write-efficient data encryption~\mbox{\cite{young2015deuce}}, multi-way wear leveling~\mbox{\cite{yu2012increasing}}, or  randomized address mapping~\mbox{\cite{seong2010security}}. 
These techniques can be readily deployed in {\optmemory}.

While several approaches discussed above address EPCM limitations, EPCM is not yet a viable alternative for DRAM. 
In Table~\ref{tab:relatedefforts}, we see that optical control of PCMs combined with silicon-photonic links significantly improves performance and lowers energy, without using any of the complementary methods provided in prior work.

\subsection{Silicon-Photonic Links and \optmemory Cells}
Silicon-photonic links have enabled high bandwidth-density and low-energy communication between processor and memory~\cite{sun2015single,batten2009building, beamer2009re, demir2014galaxy, thonnart2020popstar, narayan2020system, Batten2012Jetcas, TeraPHY, 9256327}. 
To provide high DRAM internal bandwidth, Beamer {\it et al.}~\cite{beamer2009re} proposed a joint silicon-photonic link and electro-photonic DRAM design. 
However, the O-E-O conversion in DRAM adds to the latency.
Optical control of memory cells can avoid this O-E-O conversion and enable signals in the silicon-photonic links to directly access the cells and deliver higher memory throughput.

Several recent efforts have prototyped GST-basd PCM cells with optical control.
Rios {\it et al.} demonstrate the optical control of multi-bit GST-based PCMs with fast readout and low switching energies~\cite{rios2015integrated}.
Zhang {\it et al.}~\cite{zhang2018all} present an approach to selectively couple optical signals from MRR to GST.
Feldman \mbox{\it et al.}~\mbox{\cite{feldmann2017calculating, feldmann2019integrated}} design a prototype of a monolithic {\optmemory} array based on waveguide crossing but not a comprehensive memory microarchitecture and access protocol to interface with the processor.
Subsequent efforts demonstrate higher bit density per GST~\cite{li2019fast}, in-memory computing on PCM cells using optical signals~\cite{rios2019memory}, basic arithmetic operations in \optmemory~\cite{feldmann2017calculating,feldmann2019integrated}, and a behavioral model for neuromorphic computing~\cite{carrillo2019behavioral}.
\textbf{We are the first to propose a comprehensive \optmemory microarchitecture with custom read/write access protocols, and design an \eoe unit to interface the \optmemory array with the processor.}
\section{Conclusion}
\label{sec:conclusion}

EPCM systems suffer from long write latencies and high write energies, yielding poor performance and high energy consumption for data-intensive applications.
In contrast, \optmemory technology provides the opportunity to design high-performance and low-energy memory systems due to its higher MLC capacity and the direct cell access via high-bandwidth-density and low-latency silicon-photonic links.
Adapting the current EPCM design architecture for \optmemory systems, however, raises major issues in terms of latency, energy and thermal concerns, thereby rendering such a design impractical.
We are the first to architect a complete memory system, \opcmsystem, which consists of an \optmemory array microarchitecture, a read/write access protocol tailored for \optmemory technology, and an \eoe unit that interfaces the \optmemory array with the MC.
Our evaluations show that, compared to an EPCM system, our proposed \opcmsystem system provides $2.09\times$ higher read throughput and $2.15\times$ higher write throughput, thereby reducing the execution time by $2.14\times$, read energy by $1.24\times$, and write energy by $4.06\times$.

We show that \opcmsystem designed with state-of-the-art technology provides similar performance and energy as DDR4. This is a significant finding as future higher-density \optmemory cells are expected to provide better performance.
Our promising first version of an \opcmsystem architecture opens doors for new architecture-level, circuit-level, and system-level methods to enable practical integration of \optmemory-based main memory in future computing systems.
Moreover, the high-throughput and scalable \optmemory technology ushers in interesting research opportunities in persistent memory, in-memory computing, and accelerator-specific memory designs. 



\section*{Acknowledgement}
\label{sec:acknowledgement}
\small{This work was funded partly by NSF CCF-1716352. }


\bibliographystyle{IEEEtranS}
\bibliography{opcm-micro-2021}

\end{document}